\documentclass[apj,twocolumn]{emulateapj}

\usepackage{natbib}
\usepackage{appendix}
\usepackage{amsmath}
\usepackage{amssymb}
\usepackage{graphicx}
\usepackage{verbatim}
\usepackage{lscape}
\usepackage{longtable}
\usepackage{hyperref}

\slugcomment{Submitted to ApJ on Aug 31 2016, Revised version resubmitted on Apr 11th 2017}

\shorttitle{Calibrating a Cluster Mass-Richness Relation at $0.4 < z <2.0$}
\shortauthors{Rettura et al.}

\begin{document}

%% LaTeX will automatically break titles if they run longer than
%% one line. However, you may use \\ to force a line break if
%% you desire.

\title{Mass-Richness relations for X-ray and SZE-selected clusters at $0.4 < z <2.0$ as seen by {\it Spitzer} at 4.5 $\mu$m}

%% Use \author, \affil, and the \and command to format
%% author and affiliation information.
%% Note that \email has replaced the old \authoremail command
%% from AASTeX v4.0. You can use \email to mark an email address
%% anywhere in the paper, not just in the front matter.
%% As in the title, use \\ to force line breaks.

\author{A. Rettura\altaffilmark{1}, R. Chary\altaffilmark{1}, J. Krick\altaffilmark{1}, S. Ettori\altaffilmark{2}}

\altaffiltext{1}{IPAC, Caltech, KS 314-6, Pasadena, CA 91125, USA}
\altaffiltext{2}{INAF, Osservatorio Astronomico di Bologna, via Ranzani 1, 40127, Bologna, Italy}

%\author{R. J. Hanisch\altaffilmark{5}}
%\affil{Space Telescope Science Institute, Baltimore, MD 21218}

%% Notice that each of these authors has alternate affiliations, which
%% are identified by the \altaffilmark after each name.  Specify alternate
%% affiliation information with \altaffiltext, with one command per each
%% affiliation.

%\altaffiltext{1}{Visiting Astronomer, Cerro Tololo Inter-American Observatory.
%CTIO is operated by AURA, Inc.\ under contract to the National Science
%Foundation.}
%\altaffiltext{2}{Society of Fellows, Harvard University.}
%\altaffiltext{3}{present address: Center for Astrophysics%    60 Garden Street, Cambridge, MA 02138}
%\altaffiltext{4}{Visiting Programmer, Space Telescope Science Institute}
%\altaffiltext{5}{Patron, Alonso's Bar and Grill}

%% Mark off your abstract in the ``abstract'' environment. In the manuscript
%% style, abstract will output a Received/Accepted line after the
%% title and affiliation information. No date will appear since the author
%% does not have this information. The dates will be filled in by the
%% editorial office after submission.

\begin{abstract}
 
We study the mass-richness relation of 116 spectroscopically-confirmed massive clusters at $0.4 < z < 2$ by mining the {\it Spitzer} archive. 
We homogeneously measure the richness at 4.5\,$\mu$m for our cluster sample within a fixed aperture of $2^{\prime}$ radius and above a fixed brightness threshold, making appropriate corrections for both background galaxies and foreground stars. We have two subsamples, those which have a) literature X-ray luminosities and b) literature Sunyaev-Zeldovich effect masses. For the X-ray subsample we re-derive masses adopting the most recent calibrations. We then calibrate an empirical mass-richness relation for the combined sample spanning more than one decade in cluster mass and find the associated uncertainties in mass at fixed richness to be $\pm 0.25$ dex.
We study the dependance of the scatter of this relation with galaxy concentration, defined as the ratio between richness measured within an aperture  radius of 1 and 2 arcminutes. We find that at fixed aperture radius the scatter increases for clusters with higher concentrations. We study the dependance of our richness estimates with depth of the [4.5]\,$\mu$m imaging data and find that reaching a depth of at least [4.5]= 21 AB mag is sufficient to derive reasonable mass estimates. We discuss the possible extension of our method to the  mid-infrared {\it WISE} all-sky survey data, and the application of our results to the {\it Euclid} mission.
This technique makes richness-based cluster mass estimates available for large samples of clusters at very low observational cost.
 \end{abstract}

%% Keywords should appear after the \end{abstract} command. The uncommented
%% example has been keyed in ApJ style. See the instructions to authors
%% for the journal to which you are submitting your paper to determine
%% what keyword punctuation is appropriate.

%\keywords{galaxies:    galaxies:   high-redshift   ---   galaxies:
%  fundamental  parameters   ---  galaxies:  evolution   ---  galaxies:
%  formation --- galaxies: elliptical --- cosmology: observations}

\keywords{galaxies: clusters: general   --- galaxies: high-redshift ---  galaxies: statistics  ---  cosmology: observations --- cosmology: large-scale structure of universe --- infrared: galaxies}
%galaxies:  fundamental  parameters   ---  galaxies:  evolution 

%% From the front matter, we move on to the body of the paper.
%% In the first two sections, notice the use of the natbib \citep
%% and \citet commands to identify citations.  The citations are
%% tied to the reference list via symbolic KEYs. The KEY corresponds
%% to the KEY in the \bibitem in the reference list below. We have
%% chosen the first three characters of the first author's name plus
%% the last two numeral of the year of publication as our KEY for
%% each reference.

%% Authors who wish to have the most important objects in their paper
%% linked in the electronic edition to a data center may do so by tagging
%% their objects with \objectname{} or \object{}.  Each macro takes the
%% object name as its required argument. The optional, square-bracket 
%% argument should be used in cases where the data center identification
%% differs from what is to be printed in the paper.  The text appearing 
%% in curly braces is what will appear in print in the published paper. 
%% If the object name is recognized by the data centers, it will be linked
%% in the electronic edition to the object data available at the data centers  
%%%% Note that for sources with brackets in their names, e.g. [WEG2004] 14h-090,
%% the brackets must be escaped with backslashes when used in the first
%% square-bracket argument, for instance, \object[\[WEG2004\] 14h-090]{90}).
%%  Otherwise, LaTeX will issue an error. 

\section{Introduction}

Clusters of galaxies are the largest and most massive gravitationally bound systems in the Universe. 
Clusters of galaxies are considered, at the same time, unique astrophysical laboratories and powerful cosmological probes \citep[e.g.,][]{White93, Bartlett94, Via99, Borgani01, Vikhlinin09, Rozo10, Mantz10, Allen11,Benson13, Bocquet15}. 
Clusters grow from the highest density peaks in the early universe and thus their mass function is a tracer of the underlying cosmology \citep[e.g.,][]{Press74, Bahcall93, Gonzalez 12}. 
Due to the steep dependance between number density and mass in the dark matter halo mass function, deriving cluster mass accurately is of paramount importance and large observational efforts have been devoted to this goal over the past three decades.

Different indirect methods, each of them leveraging unique observables of these systems, have been developed in the literature in order to weigh the most massive structures in the universe. These 
are:
(i) measuring the richness of a cluster, i.e. counting the number of galaxies associated with that cluster within a given radius \citep[e.g.,][]{Abell58, Zwicky68, Carlberg96, Yee99, Yee03, Rozo09, Rykoff12, Rykoff14, Andreon14, Andreon15, Andreon16, Saro15}. (ii) measuring the radial velocities of the cluster members, which yields the velocity dispersion of a cluster and can be used to derive the cluster's mass from the virial theorem, under the assumption that the structure is virialized \citep[e.g.,][]{Girardi96, Mercurio03, Demarco05, Demarco07}. (iii) measuring the intensity of the hot X-ray emitting
intracluster medium, if this gas is in hydrostatic equilibrium and by factoring in its density and temperature distribution \citep[e.g.,][]{Gioia90, Vikhlinin98, Bohringer00, Pacaud07, Suhada12, Ettori13}. (iv) measuring the inverse-Compton scatter of cosmic microwave background (CMB) photons off the energetic electrons in the hot intracluster gas. The resultant characteristic spectral distortion to the CMB is known as the Sunyaev-Zel'dovich effect \citep[SZE,][]{Sunyaev72, Staniszewski09, Hasselfield13, Planck15, Bleem15}. (v) By measuring the coherent distortion that weak gravitational lensing produces on background galaxies, which has the advantage that it does not need prior knowledge on the baryon fraction of the cluster or its dynamical state \citep[e.g.,][]{Bartelmann01,Hoekstra07, Mahdavi08, High12, Hoekstra12, vonderLinden14, Umetsu14, Sereno15}.

While there are large cluster samples selected from optical and near infrared photometric surveys up to $z<1.5$ \citep[e.g.,][]{Gladders00, Koester07, Menanteau10, Hao10, Brodwin11, Wen12, Rykoff14, Ascaso14, Bleem15a}, in recent years mid infrared photometric surveys with {\it Spitzer} have extended the landscape. 
The Infrared Array Camera \citep[IRAC,][]{Fazio04} onboard the {\it Spitzer} Space Telescope has proven to be a sensitive tool for studying galaxy clusters. Ongoing {\it Spitzer} wide-area surveys are proving effective at identifying large samples of galaxy clusters down to low masses at 1.5$<$ z$ < $2 \citep[e.g.,SDWFS, SWIRE, CARLA, SSDF;][]{Eisenhardt08,Papovich08,Wilson09,Demarco10,Galametz10,Stanford12,Zeimann12,Brodwin13,Galametz13,Muzzin13,Wylezalek13,Rettura14}, where current X-ray and SZE observations are restricted to only the most massive systems at these redshifts \citep{Brodwin11, Muzzin13}. 

Even larger samples of clusters at $0.4<z<2.0$ will soon be available from upcoming  and planned  large scale surveys like the Dark Energy Survey (DES, The Dark Energy Survey Collaboration 2005), KiDS \citep{deJong13}, {\it Euclid}
\citep{Laureijs11}, LSST (LSST Dark Energy Science Collaboration 2012), and {\it WFIRST}. However, until the next generation SZE instrumentation (e.g., ACTpol, SPTpol, SPT3G - in any case only covering the Southern sky) or next generation X-ray telescopes (e.g., eRosita, Athena) become available, measuring the masses of the bulk of the high redshift clusters at 0.4$<$z$\lesssim$2 remains challenging. 

In order to provide an efficient and reliable mass proxy for high redshift clusters up to z$\sim $2, in this paper, we calibrate a richness-mass relation using archival 4.5$\mu$m data on a sample
of published X-ray and SZE-selected clusters at 0.4$<$z$<$2.0.  At these redshifts, the 4.5$\mu$m band traces rest-frame near-infrared light from the galaxies which is emitted by the high mass to light ratio stellar population. Thus, if the integrated mass function of galaxies is correlated with the cluster dark matter halo in which they reside, the near-infrared richness should provide a reasonable tracer of cluster mass.
This method of mass measurement has the advantage over the others described above because it is  purely photometric, does not require apriori knowledge of the dynamical state of the cluster, and is observationally easy to obtain. We require only  the cluster position, an approximate redshift estimate and at least 90sec-depth coverage of IRAC 4.5$\mu$m data, over a single pointing of 5$^{\prime}\times$5$^{\prime}$ field of view.

The plan of the paper is as follows. In Section 2 we describe the archival cluster sample we have adopted throughout the work and describe how the cluster masses were derived. In Section 3 we present the {\it Spitzer} photometric cataloging procedure adopted. In Section 4 we present the definition of our richness indicator and study its dependance on survey depth and aperture radius adopted. In Section 5 we calibrate the mass-richness relation for each sub-sample individually and combined. In Section 6 we discuss our results, the possibility of extending our method  to  other mid-infrared (MIR) all-sky surveys, and the implication of our findings on future wide field infrared surveys such as those that will be undertaken with {\it Euclid}. In section 7 we summarize the results.

Throughout, we adopt a $\Omega_{\Lambda} = 0.7$, $\Omega_{m} = 0.3$ and $H_{0} = 70\ \rm{km} \rm{s}^{-1} \rm{Mpc}^{-1}$ cosmology, and use magnitudes in the AB system.

\section{Sample Selection}

In the following sections we present  the cluster samples drawn from the literature and the archival {\it Spitzer} data adopted in our analysis. Our aim is to assemble a large sample of clusters with known masses and redshifts for which archival IRAC data at 4.5 $\mu$m is publicly available. We define two cluster subsamples, based on literature X-ray masses and literature SZE masses. 

\subsection{X-ray Clusters Sample}
The starting point for this sample is the Meta-catalog of X-ray detected clusters of galaxies (MCXC), a catalog of compiled properties of X-ray detected clusters of galaxies \citep[][and references therein]{Piffaretti11}. This catalog is based on the {\it ROSAT} All Sky Survey \citep[RASS,][]{Voges99} data on 1743 clusters at $0.003< z<1.261$ that have been  homogeneously evaluated within the radius, $R_{500}$, corresponding to  an overdensity of 500 times the critical density.  For each cluster, the MCXC provides redshift\footnote{typical redshift uncertainty is  $\sigma_{z}<$ 0.001 \citep[see discussion in][]{Liu15}}, coordinates, $R_{500}$, and  X-ray luminosity in the $0.1-2.4$ keV band, $L_{500, [0.1-2.4 keV]}$. Based on  the values published in \citet{Piffaretti11}, for each cluster we also derive the angular size, $\theta_{500}$= $R_{500}/DA(z)$, where $DA(z)$ is the angular diameter distance.

In order to define a richness parameter to be used as a proxy for cluster mass, $M_{500}$, we need to define the aperture radius in which galaxies should be counted and the redshift range for which this radius is still representative of the cluster $R_{500}$.  To this aim, in Fig.~\ref{theta500} we show the entire MCXC sample $\theta_{500}$ vs. redshift relation. The red horizontal line indicates the  2.5 arcmin radius of a single {\it Spitzer}/IRAC field-of-view. The red asterisks indicate the mean $\theta_{500}$ per redshift bin of $\Delta z=0.1$, the error bars are the standard deviation of the mean per redshift bin. We note that at $z>0.4$ (dot-dashed line) the average $\theta_{500}$ of the sample is included within the {\it Spitzer}/IRAC field-of-view. Therefore we adopt this lower redshift cut to the cluster samples considered in our study. There are 142 clusters in MCXC at $z>0.4$.

The most reliable X-ray masses are obtained by solving the equation of hydrostatic equilibrium, which requires measurements of the density and temperature gradients of the X-ray emitting gas \citep[see discussion in][]{Maughan07}. This is only possible for nearby bright clusters, therefore it remains a challenge for the majority of clusters detected in X-ray surveys, especially at high redshifts where
surface brightness dimming effects become significant. Thus, in most cases, cluster masses are estimated from simple properties such as X-ray luminosities ($L_{X}$) or from adopting a single global temperature ({$kT$) via the calibration of scaling relations.  

To derive an estimate of the total cluster mass, $M_{500}$,  within $R_{500}$, we adopt the most recent calibrations, in particular in their redshift evolution, of the relations between X-ray global properties and cluster total mass, as presented in \cite{Reichert11}.
\citet[][see also references therein]{Reichert11} obtained these relations by homogenizing published estimates of X-ray luminosity and total mass. These values were rescaled at different radii and overdensities by using their dependence upon the gas density which was described by a $\beta$-model \citep{Cavaliere76}. 

\citet{Reichert11} scaling relations, together with the MCXC luminosities, are then used here to run the following iterative process:
(i) an input temperature is assumed; (ii) a conversion from the MCXC $L_{500, [0.1-2.4 keV]}$ luminosity to the pseudo-bolometric (0.01--100 keV) value, $L_{500, bol}$, is derived assuming the thermal {\tt apec} model in XSPEC \citep{Arnaud96}, adopting the temperature assumed at step (i) and a metal abundance of 0.3 times the solar value; (iii) a value of the mass within an overdensity of 500 with respect to the critical density of the universe at the cluster's redshift is then calculated from Eq.~26 in \cite{Reichert11},

\begin{equation}
\frac{M_{500, X}}{10^{14} M_{\odot}}=(1.64 \pm 0.07)\cdot\bigg(\frac {L_{500, bol}}{10^{44} erg \cdot s^{-1}}\bigg)^{0.52 \pm 0.03} \cdot \bigg(\frac{H(z)}{H_{0}}\bigg)^{\alpha},
\label{lum-mass equation}
\end{equation}

where $H(z)=\sqrt{\Omega_{\Lambda} +\Omega_{m}*(1+z)^3 +\Omega_{k}*(1+z)^2}$, $\Omega_{k}=(1-\Omega_{m}-\Omega_{\Lambda})$ and $\alpha=-0.90^{+0.35}_{-0.15}$;

(iv) a new temperature is recovered from the $M$-$T$ relation (Eq.~23 in \cite{Reichert11}) and compared to the input value assumed at step (i) ; (v) the calculations are repeated if the relative difference between these two values is larger than 5 percent.

We consider also a correction on the given luminosity due to the change in the initial $R_{500}$. 
This correction, typically few percent, is obtained as described in \cite{Piffaretti11}, by evaluating the relative change of the square of the gas density profile integrated over the cylinder with dimension of $r=R_{500}$ and height of $2 \times 5 \, R_{500}$.

\begin{figure}[htb]
\epsscale{1.0}
\includegraphics[scale=.35]{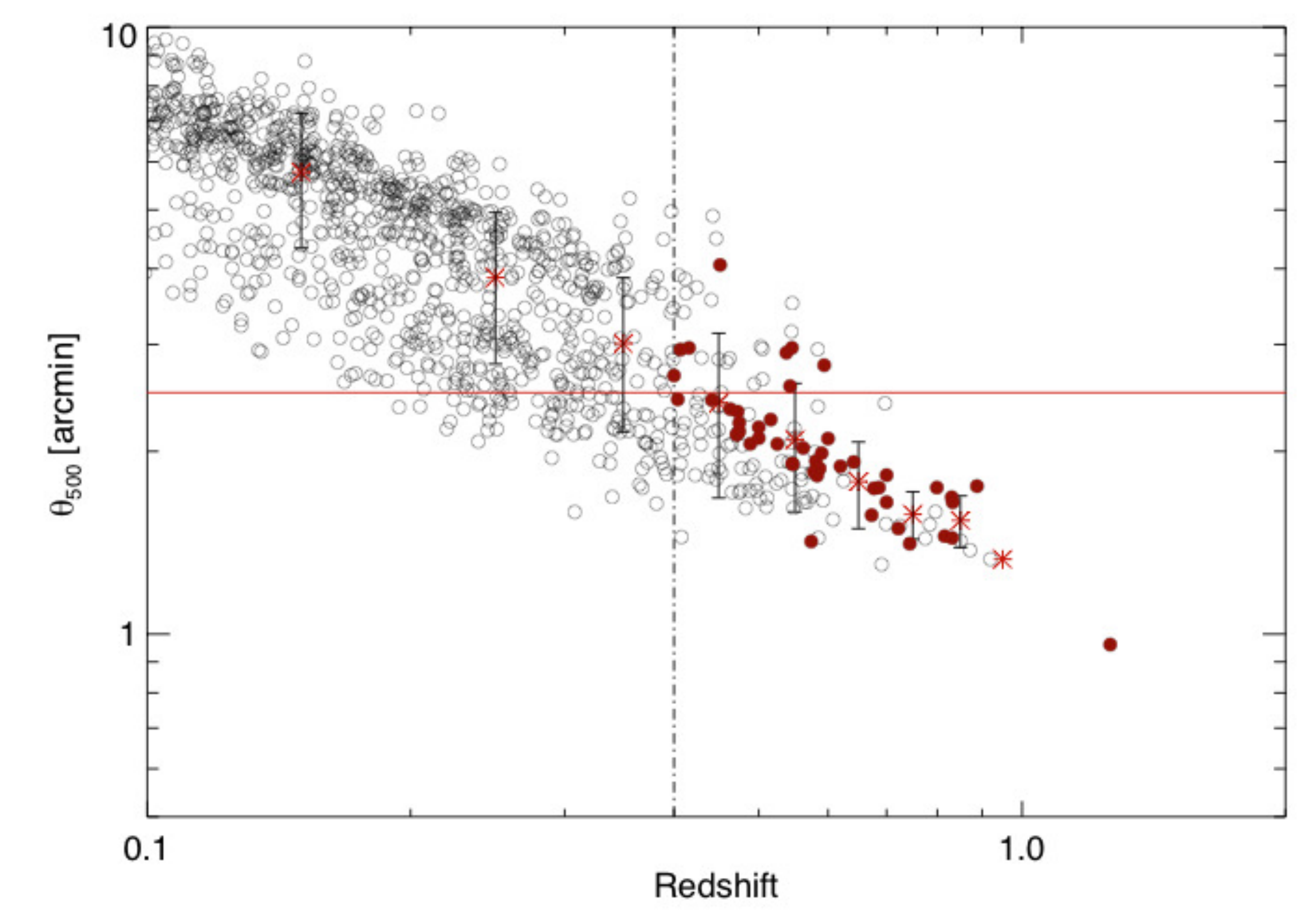}
\caption{$\theta_{500}$ as a function of redshift for the entire MCXC sample of X-ray clusters. The horizontal red line indicates half the typical field of view of a single {\it Spitzer}/IRAC pointing. At $z>0.4$ (dot-dashed line) the average $\theta_{500}$ of the sample is included in the {\it Spitzer}/IRAC field-of-view. Asterisks are average values in bins of redshift of size $\Delta z =0.1$. The final X-ray sample analyzed in this study is indicated with solid red circles.
}
\label{theta500}
\end{figure}

\begin{figure}[htb]
\epsscale{1.0}
\includegraphics[scale=.3]{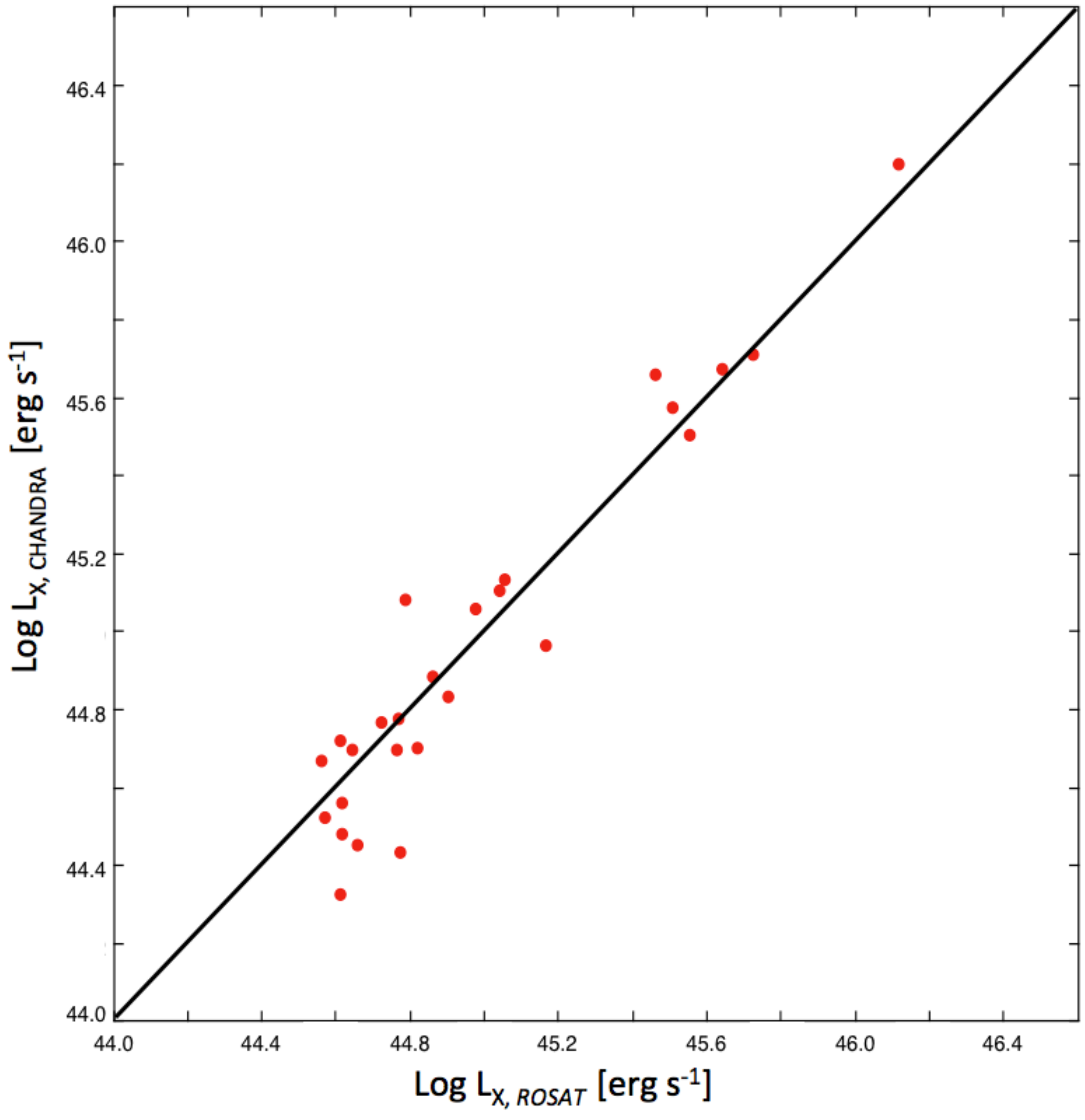}
\caption{The {\it ROSAT}-based bolometric luminosities derived in this work are plotted against those measured with {\it Chandra} by \citet{Maughan12} for a sub-sample of 26 clusters in common.}
\label{Maughan}
\end{figure}

As a consistency test, for a subsample of common clusters, we can also compare the bolometric luminosities we have obtained with those independently derived by  \citet{Maughan12}.
\citet{Maughan12} have used a sample of 115 galaxy clusters at $0.1<z<1.3$ observed with {\it Chandra} to investigate the relation between X-ray bolometric luminosity and $Y_X$ (the product of gas mass and temperature) and found a tight $L_X-Y_X$ relation  \citep{Maughan07}. They also demonstrate that cluster masses can be reliably estimated from simple luminosity measurements in low quality data where direct masses, or measurements of $Y_X$, are not possible.

There are 26 clusters  in common between our  {\it ROSAT}-based sample and their {\it Chandra} sample. In Fig.~\ref{Maughan} we compare the bolometric luminosities obtained independently and find the values to be in a very good agreement.

We then searched for {\it Spitzer}/IRAC archival observations  homogeneously covering at least an area within 2.5 arcmin radius from the cluster center coordinates, and with a minimum exposure time of 90 seconds. This depth ensures that we reach at least a 5$\sigma$ sensitivity limit of 21.46 AB mag (9.4 $\mu Jy$) at 4.5 $\mu$m (see Section 3.1 for further discussion of required depth). 

These requirements result in a final X-ray-selected sample comprised of 47 galaxy clusters at $0.4< z<1.27$ (indicated by red circles in Fig.~\ref{theta500}). We note that a few large clusters (indicated by red circles above the red line  in Fig.~\ref{theta500})  have still been considered throughout this work. This is because the mean $\theta_{500}$ in those redshift bins is smaller than the IRAC field
of view. It also ensures an adequate sample size and avoids biasing our derived richness-mass
relation against large, less-concentrated clusters. 
The derived cluster mass and redshift distributions of our X-ray sample are illustrated in Fig.~\ref{sample} (blue circles and histograms).

\begin{figure}
\epsscale{1.0}
\includegraphics[scale=.475]{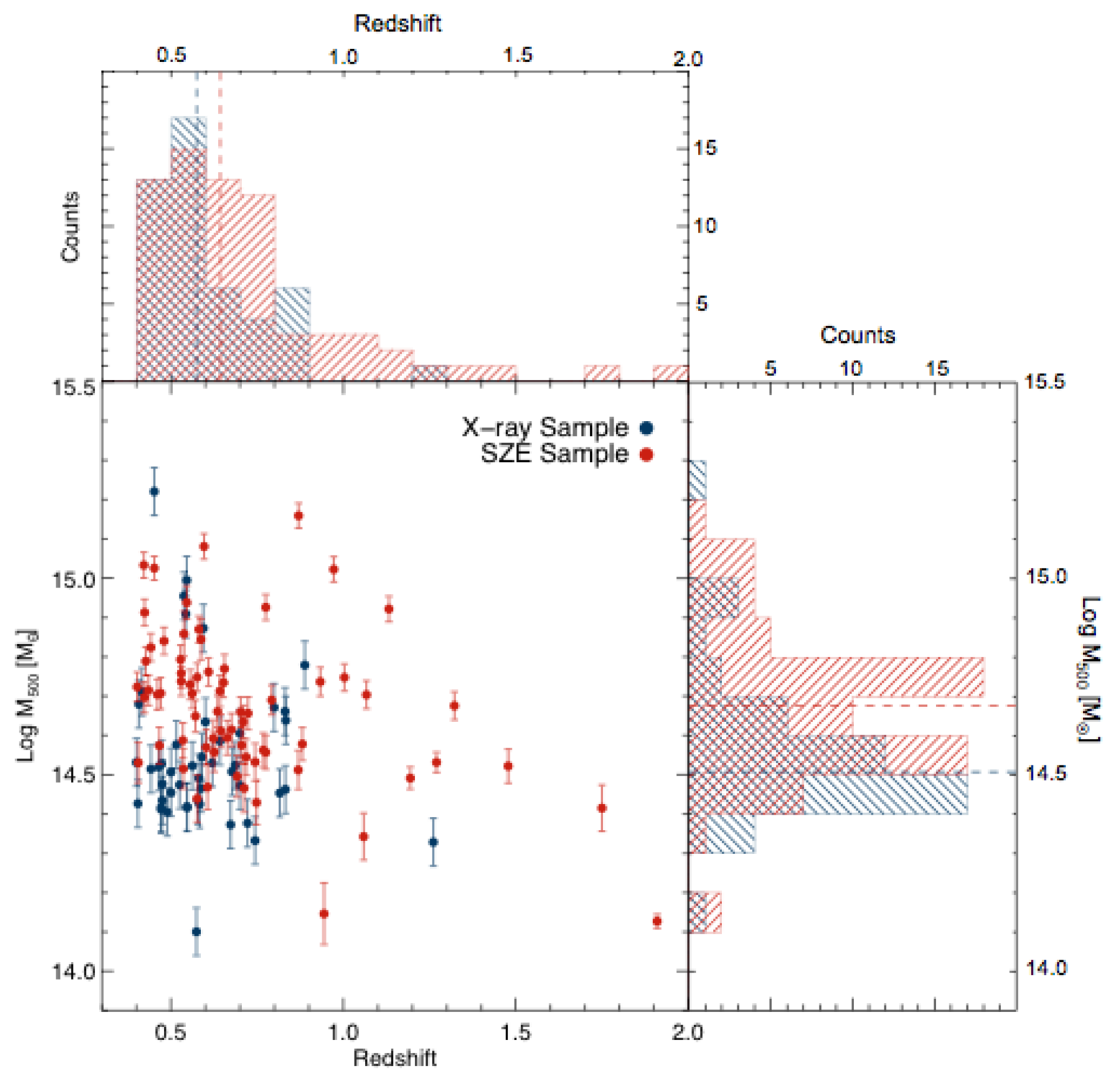}
\caption{Cluster mass and redshift distributions of the X-ray- (blue) and SZE-selected (red) cluster samples studied in this work. Both subsamples extend over similar ranges of the $M_{500} - z$ plane, and the median values are indicated by the dashed lines of the corresponding color.}
\label{sample}
\end{figure}

\subsection{SZE Clusters Sample}
Recent years have seen rapid progress of both the quality and quantity of SZE measurements, using a variety of instruments. Therefore, several programs have been launched in the past few years  with the aim of measuring total masses through the SZ effect of large samples of clusters, both for cosmology and astrophysics studies.
 {\it Spitzer}/IRAC data coverage over some of the SZ survey fields have been requested, as well as targeted SZE observations of existing MIR-selected clusters have also been obtained by various investigators.

We have therefore mined the  {\it Spitzer}/IRAC archive and drawn a heterogeneous sample of, spectroscopically confirmed, SZE selected clusters, based from a number of these programs.
Applying the same redshift and photometric coverage selection criteria illustrated in \S 2.1, the final SZE-selected sample considered in our study is comprised of 69 galaxy clusters at $0.4< z<2.0$. 

In particular, our sample is comprised of 4 clusters from the {\it Planck} Cluster Catalog \citep{Planck15},  4 clusters from the Massive Distant Clusters of {\it WISE} Survey \citep[MADCoWS,][]{Brodwin15}, 1 cluster from the IRAC Distant Cluster Survey \citep[IDCS,][]{Brodwin12},  1 cluster from the {\it XMM}-Newton Large Scale Structure Survey \citep[XLSSU,][]{Pierre11, Mantz14}, and 59 clusters from the SPT-SZ Cluster Survey \citep[SPT-SZ,][]{Bleem15},. 

Cluster masses, $M_{500, SZ}$, as reported in the aforementioned papers, are based on the spherically integrated Comptonization measurement, $Y_{500, SZ}$, obtained by either the {\it Planck} space telescope, the Combined Array for Research in Millimeter-wave Astronomy (CARMA\footnote{\url{https://www.mmarray.org}}) or the South Pole Telescope \citep[SPT;][]{Carlstrom11, Austermann12, Story13}.
Cluster mass and redshift distributions of the final SZE-selected sub-sample are illustrated in Fig.~\ref{sample} (red circles and histograms) and can be compared with the X-ray sample shown therein.

\section{{\it Spitzer} Data}

Publicly available {\it Spitzer}/IRAC data for each cluster in our sample is accessible via the {\it Spitzer Heritage Archive} (SHA).  All of the IRAC data for the X-ray-selected sample were acquired during the initial cryogenic mission, while all, but four, of the SZE sample data were acquired during the post-cryogenic {\it Warm Mission}. The {\it Warm} and {\it Cryo} missions have been put onto the same calibration scale in the SHA provided data products, so we expect no differences between the missions to be relevant to this work.

\subsection{Source Extraction}

The publicly accessible  Spitzer Enhanced Imaging Products\footnote{\url{http://irsa.ipac.caltech.edu/data/SPITZER/Enhanced/SEIP}} (SEIP) provide super mosaics (combining data from multiple programs where available) and a source list of photometry for  sources observed during the cryogenic mission of {\it Spitzer}. The SEIP includes data from the four channels of IRAC (3.6, 4.5, 5.8, 8 $\mu$m) and the 24 $\mu$m channel of the Multi-Band Imaging Photometer for Spitzer (MIPS) where available.  In addition to the {\it Spitzer} photometry, the source list also contains photometry for positional counterparts found in the AllWISE release of the Wide-Field Infrared Survey Explorer (WISE) and in the Two Micron All Sky Survey (2MASS). To ensure high reliability, strict cuts were placed on extracted sources, and some legitimate sources may appear to be missing. These sources were removed by cuts in size, compactness, blending, shape, and SNR, along with multi-band detection requirements. In most fields, the completeness of the source list is well matched to expectations for a SNR=10 cut off, as reliability is favored over completeness. However, the list may be incomplete in areas of high surface brightness and/or high source surface density.  This is most relevant for this work for objects near bright sources or the centers of clusters which may have a higher source density.

Following the recommendations in \citet{Surace04}, for our richness estimate we adopt the aperture corrected, IRAC 4.5 $\mu$m flux density measured within an aperture of diameter 3.8 arcseconds from the SEIP source list. The chosen aperture is twice the instrumental FWHM, which provides accurate photometry , with an aperture correction for a point-source already applied, which is customary for cluster studies with {\it Spitzer} in the literature  \citep[e.g.,][]{Bremer06, Rettura06}. IRAC PSF has a FWHM$\sim$2 arcsec, thus we note that a star/galaxy separation in {\it Spitzer} data, especially at faint fluxes, is not straightforward. Therefore we will describe in \S 4 how we account and correct for foreground stars in our richness estimates.
 
For the {\it Warm Mission} data a SEIP source list is not available in the {\it Spitzer} archive. However we  have adopted the same SEIP source extraction pipeline and applied it ourselves in exactly the same way as for the {\it Cryo} mission clusters.  

\begin{figure}[t!]
\epsscale{1.0}
\includegraphics[scale=.3]{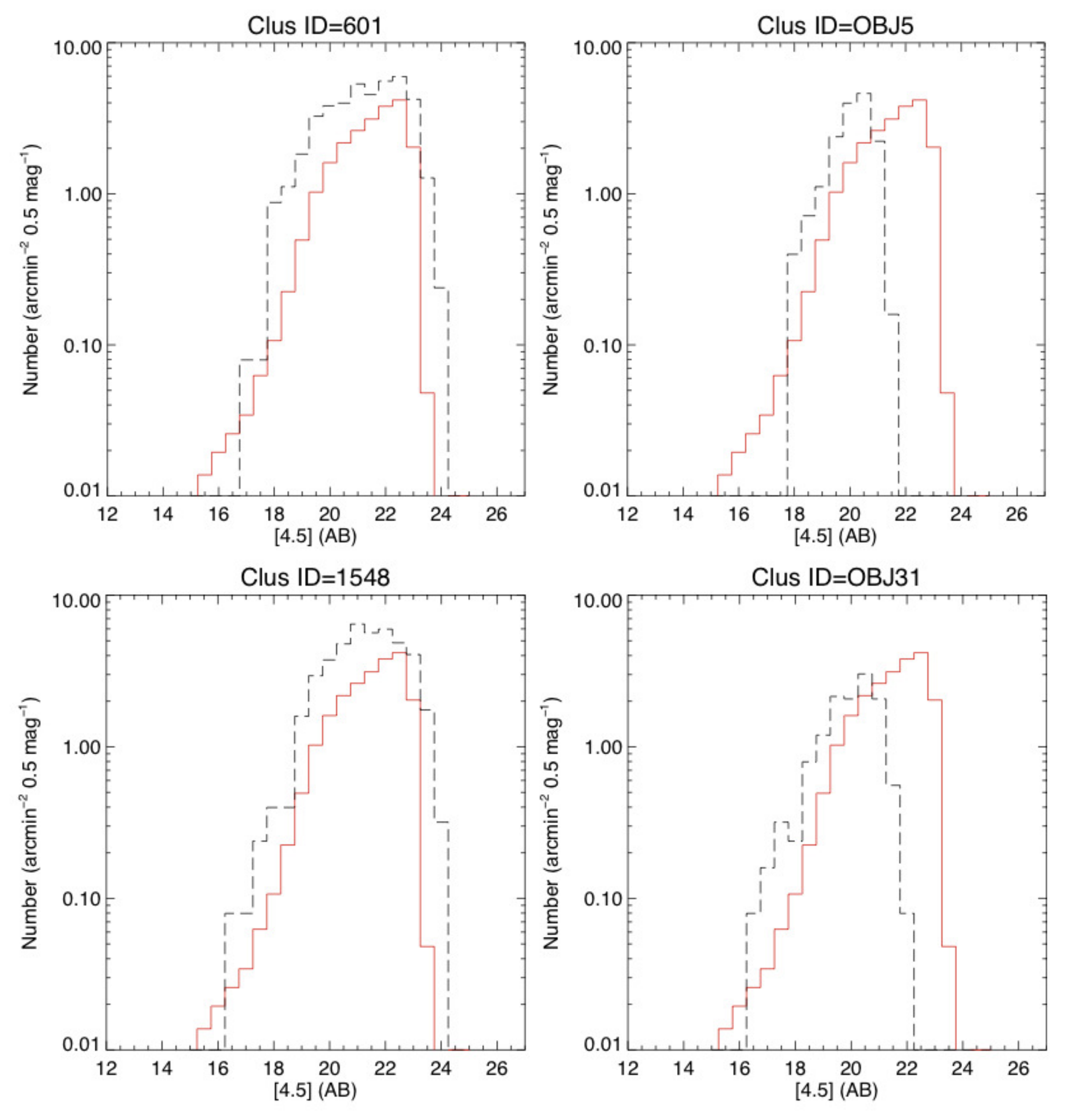}
\caption{4.5$\mu$m number counts  for four representative clusters in our sample (black dashed histogram), compared to number counts of the deeper SpUDS control field (red solid histogram). The left column shows the number counts for two X-ray  clusters while the right column shows the number counts for two SZE clusters.}
\label{numcounts}
\end{figure}

\begin{figure}[t!]
\epsscale{1.0}
\includegraphics[scale=.55]{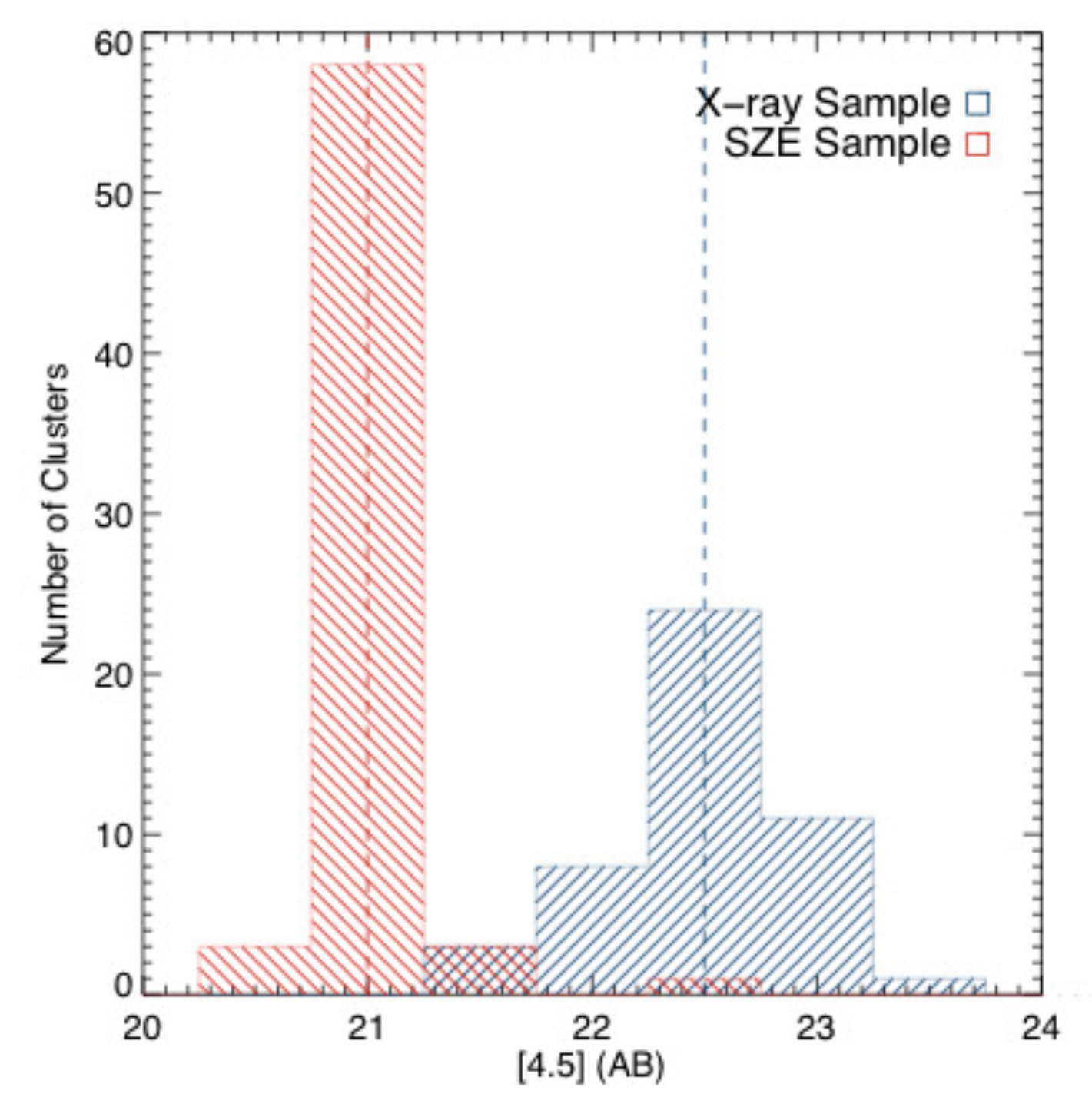}
\caption{Histogram of the 4.5$\mu$m depths reached by the archival data available for the  X-ray (blue) and the SZE (red) samples. The dashed lines indicate the median depth of each sample.}
\label{histodepth}
\end{figure}

\begin{figure*}[h!]
\epsscale{1.0}
\includegraphics[scale=.6]{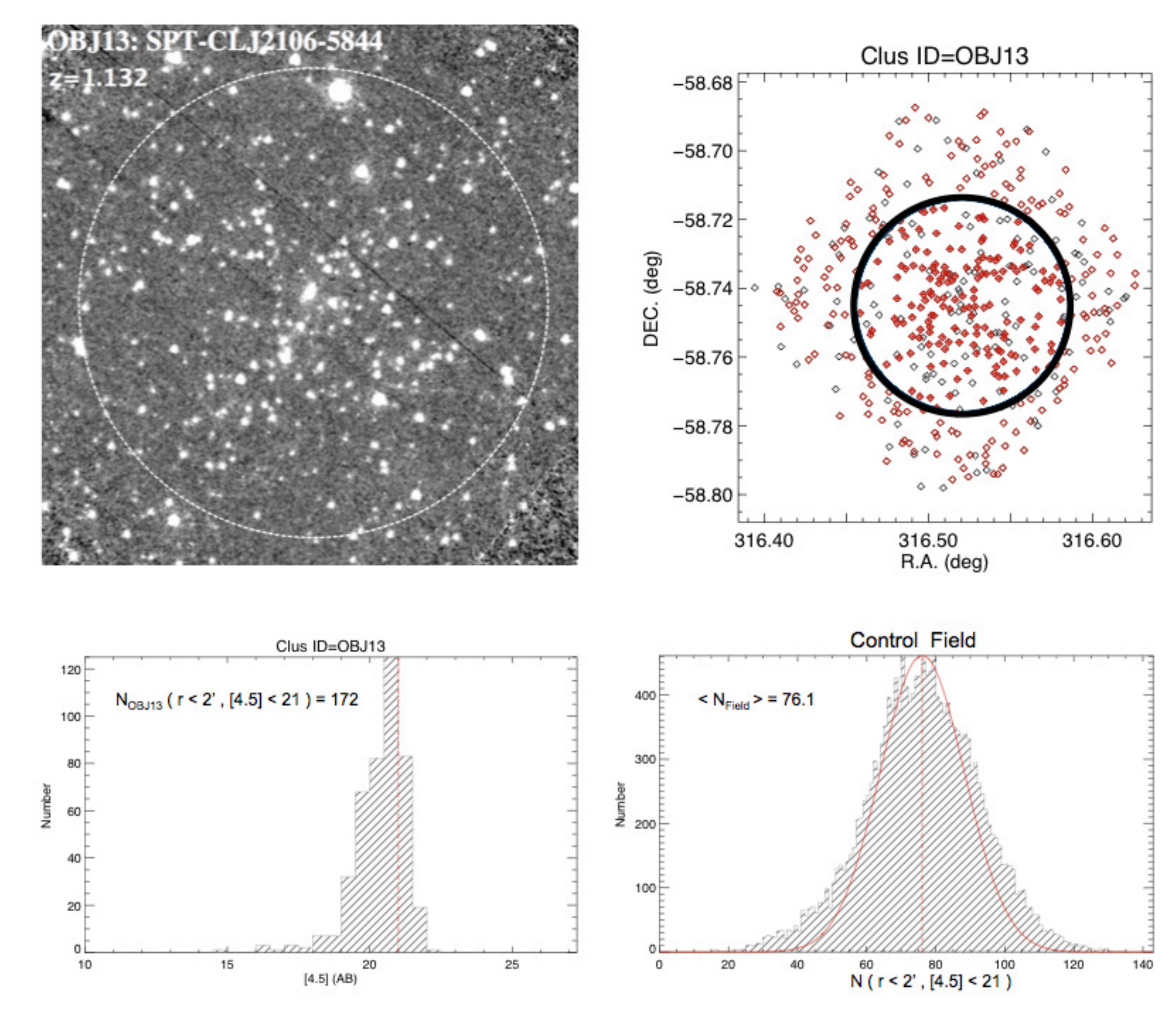}
\caption{{\it Top-left Panel:} $[4.5]$-band image of a representative cluster in our sample at $z=1.132$. The white dashed circle indicated has a radius, $r=2'$. {\it Top-right Panel:} Positions of all the sources extracted by the photometric pipeline from the 4.5$\mu$m band image of the cluster and indicated by black diamonds. Sources with magnitudes $[4.5]< 21$ AB are indicated by open red diamonds. The black circle has a radius, $r=2'$, centered on the reported cluster center. Magnitude-selected sources that are also within the circle are indicated with filled red symbols. {\it Bottom-left Panel:} [4.5] magnitude distribution of all sources in the {\it Spitzer}/IRAC image of this cluster. The  red dot-dashed line indicates the magnitude cut adopted consistently throughout this work. The number of sources,$N_{Cluster}$, brighter than the $[4.5]_{cut}$=21 AB and within 2$\arcmin$ from the cluster center is indicated.  {\it Bottom-right Panel:} Distribution of the number of sources in the control field brighter than $[4.5]= 21$ and within $r<2'$ from each source extracted in the SpUDS photometric catalog. The red line indicates a 2-$\sigma$ clipped Gaussian fit of the distribution. The red dot-dashed line indicates the mean of the Gaussian fit, $<N_{Field}>$, that is used for the source background correction throughout this work, as described in the text. Clearly, the cluster field has more than twice
as many objects within the aperture.}
\label{method}
\end{figure*}

\begin{figure}[t!]
\epsscale{1.0}
\includegraphics[scale=.31]{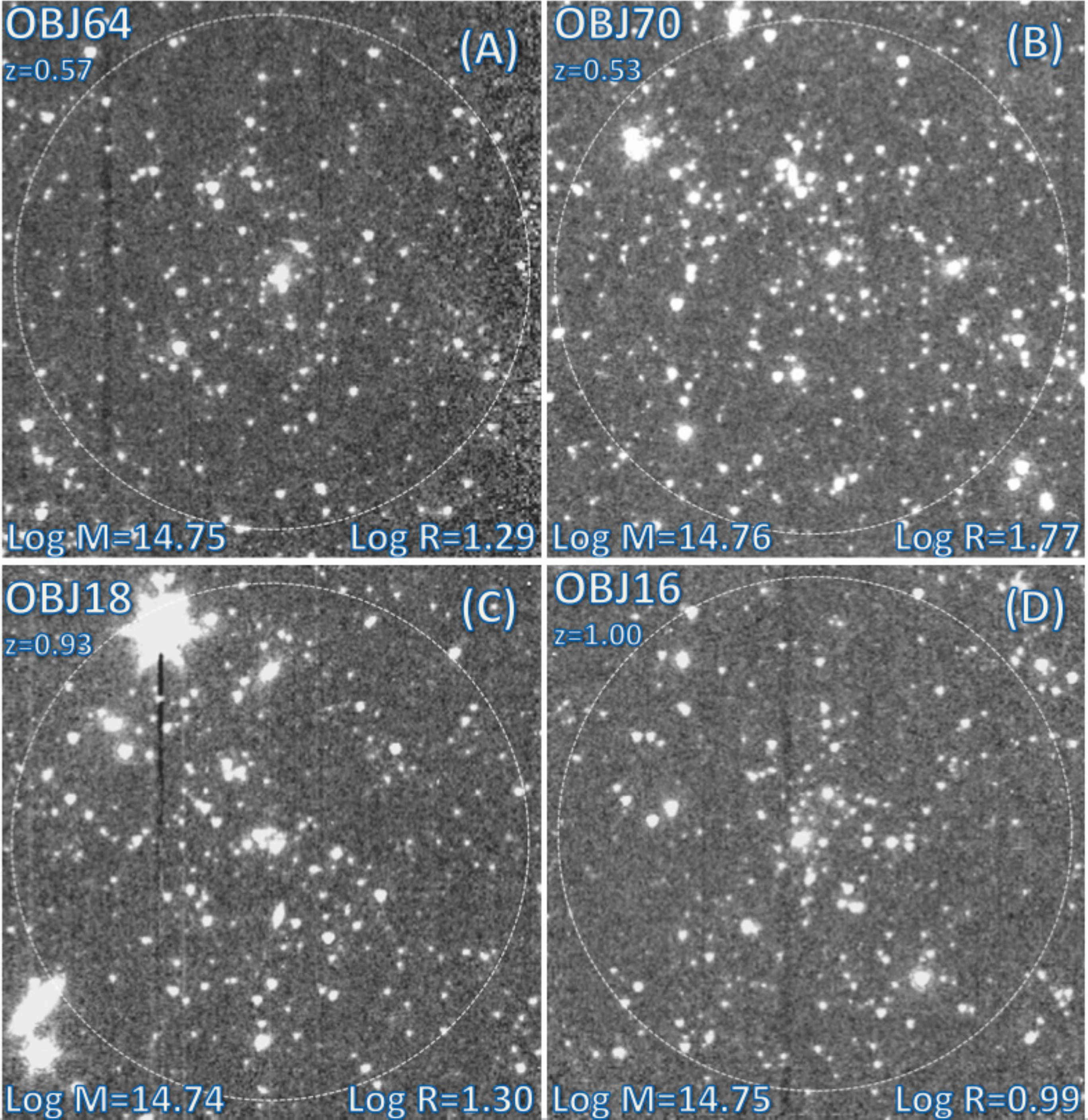}
\caption{ {\it Spitzer}/IRAC images of four clusters in our sample with similar M$_{500,SZ}$ but with different richness or redshifts measured. The white-dashed circles have a radius, $r=2'$, centered on the reported cluster. Images (A) and (B) show clusters at similar redshift, $z\sim 0.5$, while (C) and(D) are instead at  $z\sim 1$. Despite having similar redshifts and masses, (A) and (B) are found with large differences in their richness values suggesting that there may be other dependancies. Images (A) and (C) however show clusters of the same mass, found with similar richness despite being at different redshift.}
\label{Panel_method}
\end{figure}

\subsection{Surveys Depth}

As we deal with a heterogenous sample that has been observed by {\it Spitzer} at varying depths, for consistency of our analysis, we aim to be able to calibrate our method to a depth that is reached by all our archival data.

For illustration purposes, we show in Fig.~\ref{numcounts} the number counts of four representative clusters in our samples along with the number counts derived from a reference deep Spitzer legacy program, that we adopt as a control field. The {\it Spitzer} UKIDSS Ultra Deep Survey (SpUDS, PI: J. Dunlop) data used here come from a program covering $\sim$1 $deg^{2}$ in the UKIRT Infrared Deep Sky Survey, Ultra Deep Survey field \citep[UKIDSS UDS][]{Dye06}, centered at R.A.=02$^{h}$:18$^{m}$:45$^{s}$, Decl.$=-05^{\circ}$:00$^{'}$:00$^{''}$.  Note that we use the SEIP source list photometry available in the archive for our control (SpUDS) field as well.  The SpUDS survey reaches greater sensitivities than the data on the majority of our clusters, in particular for the SZE sample, as shown by examples on the right column of the panel Fig.~\ref{numcounts}.

As shown in Fig.~\ref{histodepth} for the entire sample, the IRAC coverage of our samples is not uniform. The median depth  of the SZE cluster observations reach $[4.5]=21$ AB, for instance, while the median depth for the X-ray sample reaches $[4.5]=22.5$ AB. For the sake of overall consistency of our analysis and to be able to calibrate our method to a depth that the vast majority of current and future {\it Spitzer} surveys can easily reach (with even 90 second exposure), we adopt $[4.5]_{cut}=21$ AB as the magnitude cut for all subsequent analyses. We will also further investigate the dependance of richness estimates on image depth in \S4.1.

For galaxy stellar populations formed at high redshift, a negative k-correction provide a nearly constant 4.5$\mu$m flux density over a wide redshift range. An $L_{[4.5]}^{*}$ galaxy formed at $z_{f}$ = 3 will have $[4.5] \sim 21$ (AB) at $0.4 \lesssim z \lesssim 2.0$, which is sufficiently bright that it is robustly seen in even just 90 seconds integrations with Spitzer \citep[e.g.,][]{Eisenhardt08}.

While we recognize that using the simple approach of a single  apparent magnitude cut  at $0.4<z<2.0$ would introduce a bias for optical mass-richness relationships, we note here that an infrared relation is not significantly affected because of the k-correction. Adopting \citet{Mancone10} results on the evolution with redshift of the characteristic absolute magnitude $M^{*}_{[4.5]}(z)$, we note that  at 4.5$\mu$m, in the redshift range spanned by our sample, the stellar population evolution and redshift evolution  are roughly matched, thus sampling a similar rest-frame luminosity range of the cluster galaxy population as a function of redshift. Because of cluster galaxy population evolution with redshift seen through the 4.5$\mu$m band filter at $0.4\lesssim z \lesssim 2$, our adopted apparent magnitude limit $[4.5]_{cut}$ always corresponds to a roughly similar absolute magnitude $M_{[4.5]_{cut}}\sim M^{*}_{[4.5]}(z)+1$ over this large redshift range. We find in fact that $M_{[4.5]_{cut}}$ varies between $M^{*}_{[4.5]}(z)$+0.87 (at $z\gtrsim$1.2) and $M^{*}_{[4.5]}(z)$+1.17 (at z$\sim$0.5), thus by 0.3 mag. This small variation in limit magnitude will not significantly increase the scatter in the Mass-richness relation we will derive in the next \S5. In Section 4.1, based on a sub sample of clusters for which deeper data are available, we will study the dependence of Richness estimates on survey depth and will parameterize a linear relation (Eq. ~3) to account for these effects. Accordingly, a  variation in magnitude cut by 0.3 mag will result in a variation in Richness, $\Delta R\sim6 gals \times Mpc^{-2}$, hence in logarithmic scale  $\Delta Log R\sim$ 0.05, which is very small and will not significantly increase the scatter in the derived Mass-richness relation.

\section{Derivation of {\it Spitzer} 4.5$\mu$m Richness }

\begin{figure}[t!]
\epsscale{1.0}
\includegraphics[scale=.575]{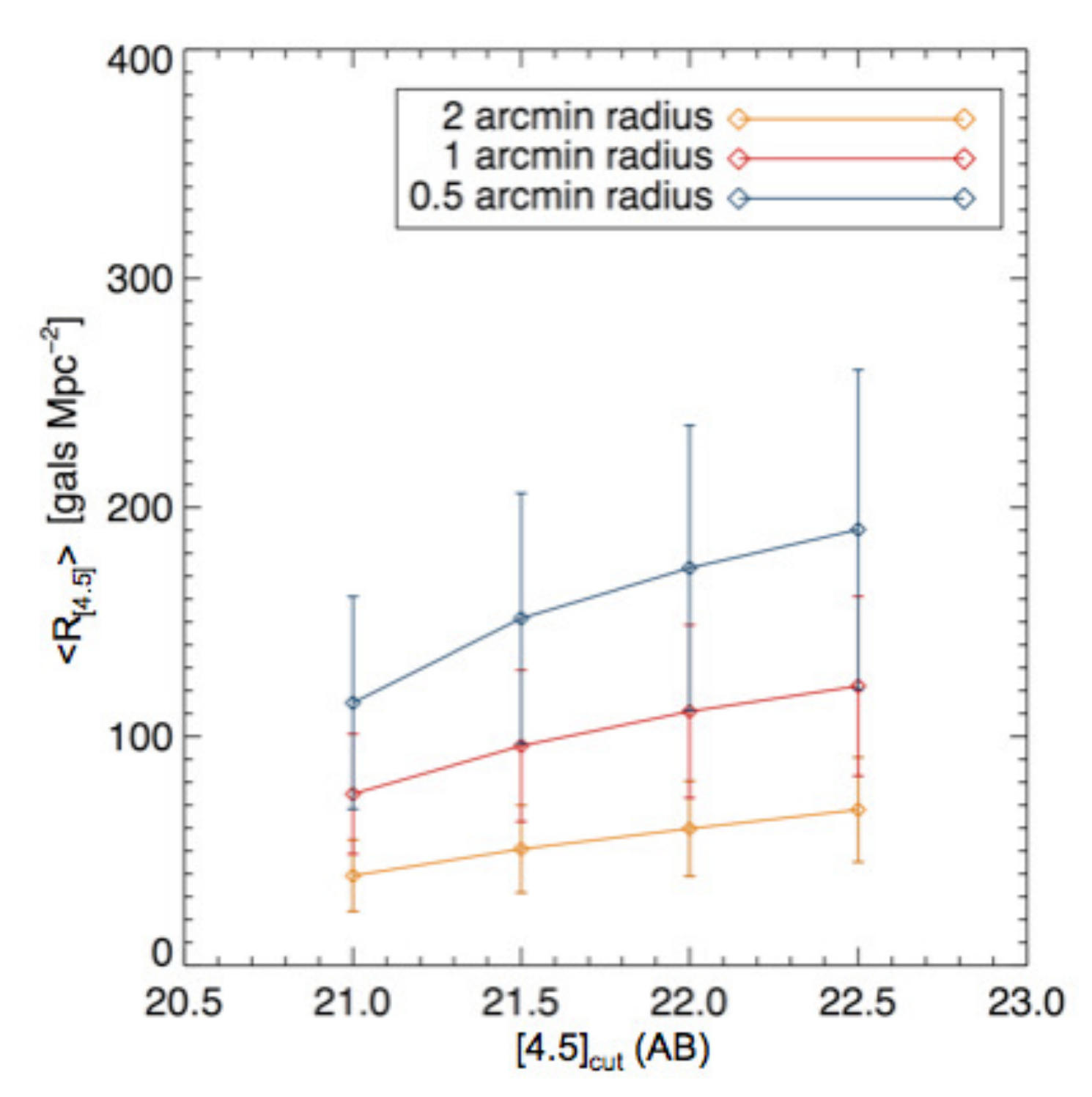}
\caption{The dependance of average richness, $<R_{[4.5]}>$  with adopted 4.5$\mu$m magnitude cuts,  $[4.5]_{cut}$, and aperture radii for the reference X-ray `deep sample'. Error bars indicate the standard deviation of the mean.}
\label{magdepth}
\end{figure}

The richness of a cluster is a measure of the surface density of galaxies associated with that cluster within a given radius. Because of the presence of background and foreground field galaxies and foreground stars, one cannot identify which source in the vicinity of a cluster belongs to the cluster. Richness is therefore a statistical measure of the galaxy population of a cluster, based on some operational definition of cluster membership and an estimate of foreground/background subtraction. 
Furthermore, as we aim to provide an efficient and inexpensive 4.5$\mu$m photometric proxy of cluster mass within $R_{500}$, we need to adopt a sufficiently large aperture radius in which galaxies should be counted that minimizes the Poisson scatter in richness, takes into account the typical $R_{500}$ of clusters at $z>0.4$ and the angular size
constraint defined by the single pointing {\it Spitzer} field-of-view. Thus we define a richness parameter, $R_{[4.5]}$, as the background-subtracted projected surface density of sources with $[4.5]<21$ AB  within 2 arcminutes from the cluster center, expressed in units of $galaxies \cdot Mpc^{-2}$. 

We first measure the number of objects in the vicinity of the cluster, $N_{Cluster}$, with $[4.5]<21$\,mag  within 2 arcminutes of the cluster center determined from the SZE or X-ray data (bottom-left panel of Fig.~\ref{method}). In order to estimate the number of background sources (stars and galaxies) to subtract, we use the SpUDS survey to derive a mean blank field surface density of sources
above the same magnitude limit. 
To estimate this, we measure the number of sources above the magnitude limit within an aperture radius of 2 arcminutes from each source with $[4.5]<21$ in the SEIP photometric catalog of the  SpUDS field.  We then fit a Gaussian to the distribution, iteratively clipping at 2$\sigma$ (see bottom-right panel of Fig.~\ref{method}). The resulting mean of the distribution, $<N_{Field}>=76~gals$ is then subtracted from $N_{Cluster}$. 

This method of background subtraction however assumes that the stellar density in the SpUDS field is the same as that in the cluster field which need not be true due to the structure of our Galaxy.
As we deal with an all sky, archival sample of clusters, we correct for the variation of the foreground star counts with Galactic latitude.  Using the \citet{Wainscoat92} mode predictions\footnote{web tool available at: \url{http://irsa.ipac.caltech.edu/applications/BackgroundModel/}}  for the IR point-source sky, we can estimate the number of stars with $[4.5]<21$  within 2 arcminutes from the center of each cluster in our sample, $N_{S}$, and compare it to the average value for the SpUDS field, $N_{S, Field}=9.4$. Thus we can correct our richness estimate at the location of each cluster for the difference in star counts by subtracting the difference between these numbers:
\begin{equation}
R_{[4.5]}= N_{Cluster} - <N_{Field}> - (N_{S} - N_{S, Field}) 
\label{rich corr}
\end{equation}
$N_{Cluster}$ and $N_{S}$ for each cluster in our sample is listed in Table 1 and 2.

To test the fidelity of the calibrated model of the Galaxy adopted here, we have also compared \citet{Wainscoat92}  predictions with the ones from a more recent model of the Galaxy, TRILEGAL \citep{Girardi12} .
At each of the 116 cluster positions, TRILEGAL has been run 10 times with varying input parameters (IMFs, extinction laws, model of the thin/thick disk, halo and bulge model) to output the mean and stdev values for the number of stars within 2 arcminutes. We find the results of the two models in remarkably good agreement. At the coordinates of our sample clusters, we find the median difference between the outputs of the two models to be only $\sim$ 1.5 stars in a 2 arcminute radius. The mean difference of the two models is found to be $\sim$ 4.3 stars in a 2 arcminute radius. This difference is two order of magnitude smaller than the typical total source counts, $N_{Cluster}$, at the location of the clusters (see Table~1 and 2) hence negligible with respect to the typical errors (Poissonian statistics) reported here.
 
 Finally we normalize for the surface area subtended by the 2$\arcmin$ radius aperture
at the redshift of each cluster  and express $R_{[4.5]}$ in units of $gals \cdot Mpc^{-2}$ throughout the paper (unless specified).  We note that since projected areas evolve slowly with redshift, in particular at high redshift, our method is suitable also for clusters for which only a photometric redshift is available. For instance, for a $z_{phot}$=1.0 cluster, even a large uncertainty in redshift of $\Delta z = \pm 0.1$ would only result in a variation of area of just $\sim5\%$,  implying a small variation of the inferred $R_{[4.5]}$.

The derived richness values for our sample of clusters are listed in Table 1 and 2. Richness uncertainties account for Poisson fluctuations in background counts and cluster counts as well as the uncertainty in the mean background counts shown in Figure \ref{method}. 

We note that we do not adopt a color criterion in our richness definition. The $[3.6]-[4.5]$ color is known to be degenerate with redshift at $z \lesssim 1.3$, but can be used as an effective redshift indicator \citep[e.g.,][]{Papovich08,Muzzin13,Wylezalek13,Rettura14} at $z > 1.3$. The method takes advantage of the fact that the $[3.6]-[4.5]$ color is a linear function of redshift between $1.3 \lesssim z \lesssim 1.5$ and at $z \gtrsim 1.5$ the color reaches a plateau out to $z \sim 3$ (see also left panel of Fig.~\ref{models}). While an IRAC color cut $[3.6]-[4.5] > -0.1$ (AB) is effective at identifying galaxies at $z >$ 1.3, due to the color degeneracy at lower redshifts, having at least one shallow optical band in addition, would be required for alleviating contamination from foreground interlopers at $z <0.3$ (see discussion in \citet{Muzzin13}).
Since optical data are unavailable for the large part of our archival sample and $>90$ \% of our sample is  comprised of cluster galaxies at  $z < 1.3$ we do not include a color cut in our definition of richness. We also note that by measuring richness at 4.5\,$\mu$m, corresponding to rest-frame near-infrared bands at the redshifts spanned by our sample, we are tracing the masses of galaxies
better than optical richness estimates because stellar mass-to-light ratios show less scatter in the NIR than in the optical \citep[e.g.,][]{Bell01}.

\subsection{Dependance of Richness on survey depth and aperture radius}

In order to investigate the effect of the chosen aperture radius and depth of the IRAC $4.5 \mu$m data on richness estimates, we performed a series of tests on a subsample of clusters where the IRAC data is deep enough to allow us to measure richness values at different sensitivity levels. 
As shown in Fig.~\ref{histodepth}, the X-ray sample contains a `deep' subsample of 36 clusters for which their depth is  $\geq 22.5$ mag AB. We measure the average (and standard deviation of the mean) richness of this sample down to various depths, $[4.5]_{cut}$, ranging $21 < [4.5] < 22.5$, and with different aperture radii, $0.5'<r<2'$. As shown in Fig.~\ref{magdepth}, richness increases with increasing magnitude cut adopted which is not surprising since there are typically more galaxies at fainter luminosities for a canonical luminosity function.  This test validates the importance of adopting a uniform magnitude cut while dealing with a heterogeneous, archival sample. We also note that the slope and standard deviation of richness is much smaller for the larger, adopted, 2 arcminutes radius aperture
than for the smaller apertures.
For the adopted  2 arcminutes aperture radius, the dependance of the mean richness, $ < R_{[4.5]} > $, with magnitude cut adopted,  $[4.5]_{cut}$, is best-fitted with the linear relation

\begin{equation}
\frac{<R_{[4.5]}>}{galaxies \cdot Mpc^{-2}} = (-358.71) +  (18.97) \times \frac{[4.5]_{cut}}{AB\hspace{0.1cm}mag}.
\label{richeness depth equation}
\end{equation}

This relation allows us to quantify the expected increase in average richness value, at increasing depths of the observations, due simply to an intrinsic photometric effect, not  due to cluster-to-cluster variations or dynamical state. By means of extrapolations, this relation could be also used to predict the expected average richness value for samples of clusters at similar redshifts with upcoming,  wide-area, infrared surveys (see discussion in \S 6.4).

\begin{figure}[t!]
\epsscale{1.0}
\includegraphics[angle=-90,scale=.52]{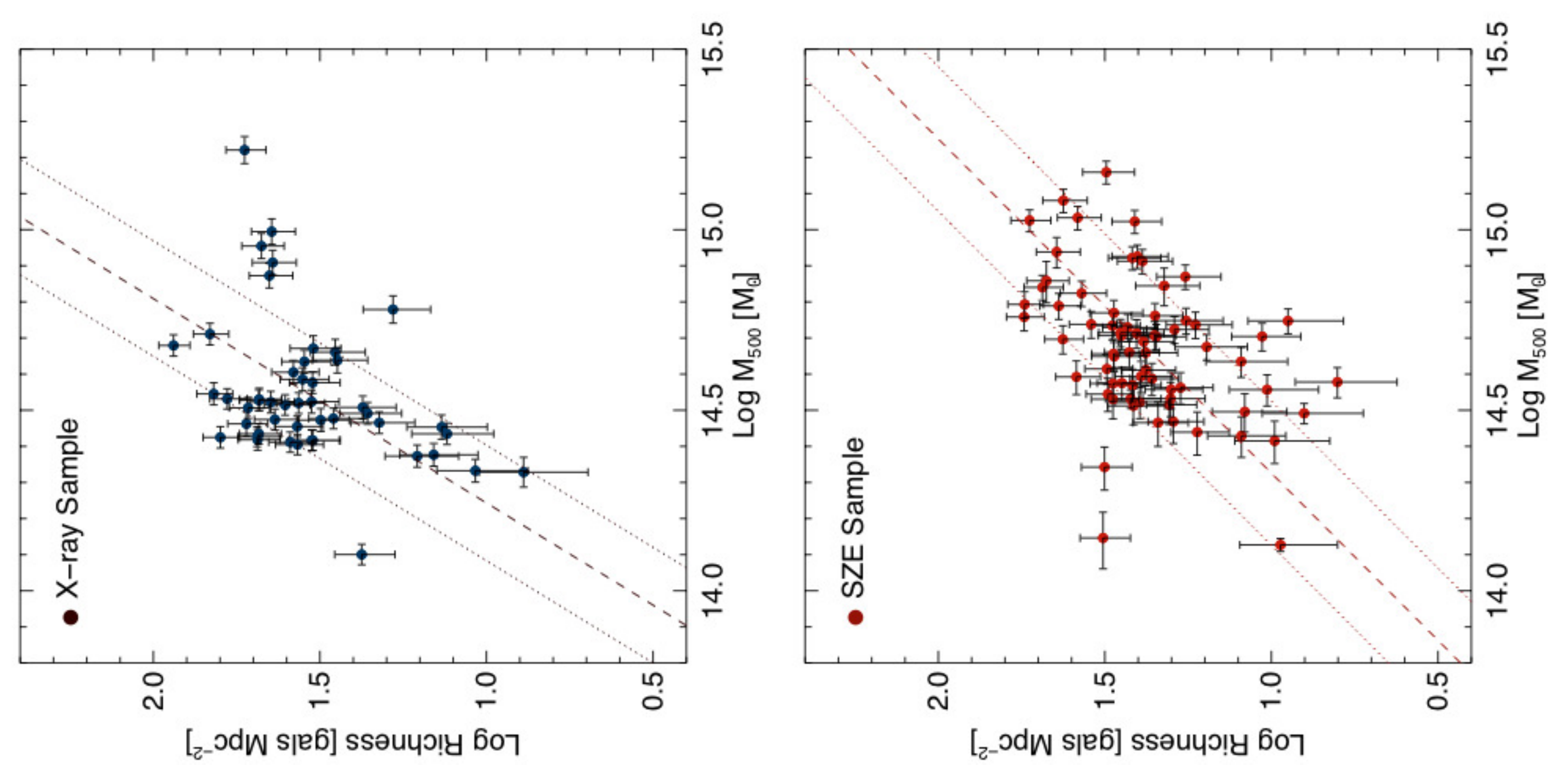}
\caption{The 4.5$\mu$m richness-mass relation for a sample of 47 X-ray selected clusters (top panel) and 69 SZE-selected clusters (bottom panel) at $0.4<z<2.0$. The dashed lines correspond to the best straight-line fits to data with errors in both coordinates for each sample respectively.  The dotted lines indicate the 68.3$\%$ confidence regions of each fit.}
\label{MassRichnessXvsSZ}
\end{figure}

\begin{figure}[h!]
\epsscale{1.0}
\includegraphics[scale=.285]{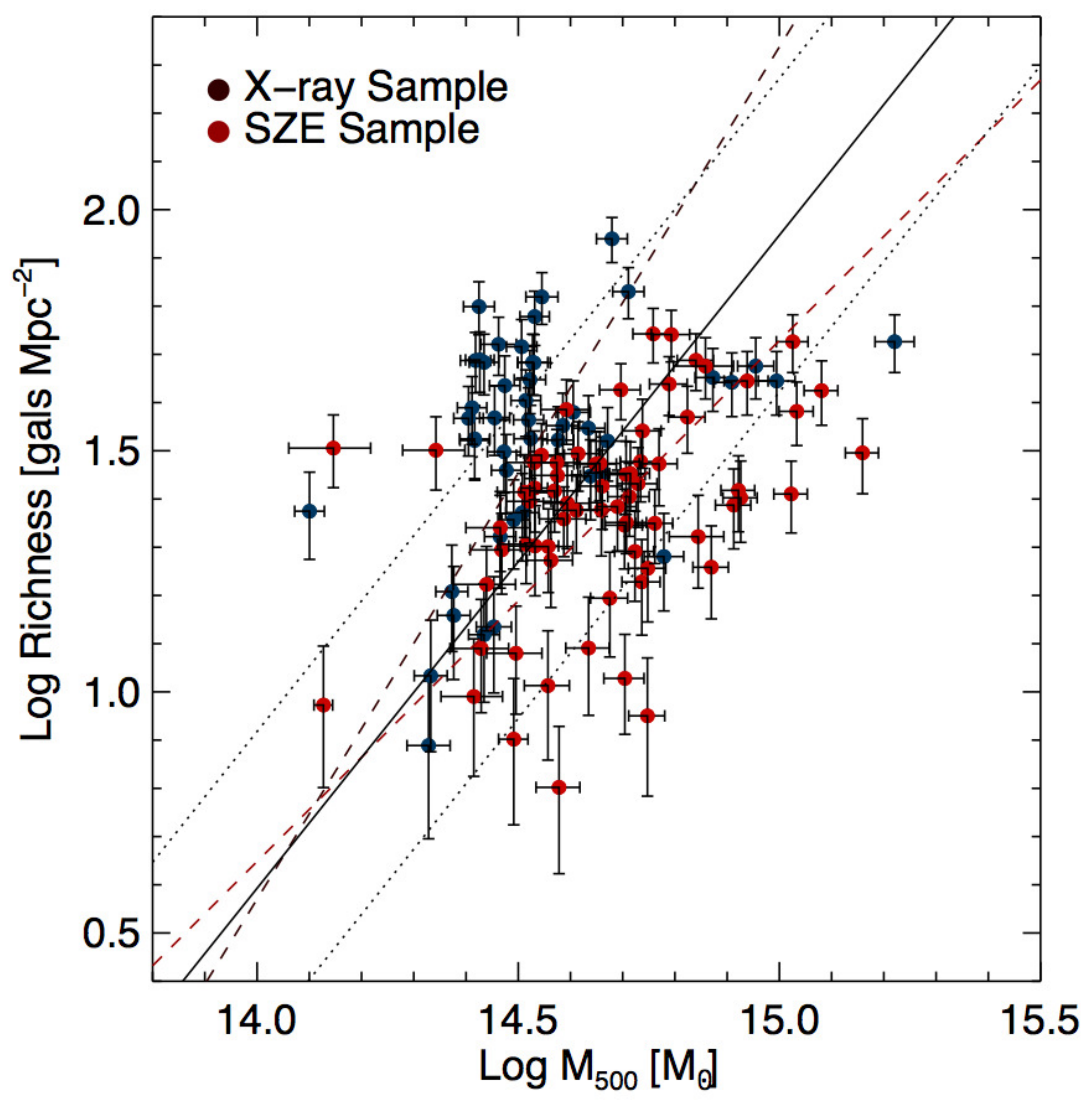}
\caption{The 4.5$\mu$m richness-mass relation for a sample of 47 X-ray selected clusters (blue circles) and 69 SZE-selected clusters (red circles) at $0.4<z<2.0$. The blue and red dashed lines correspond to the best straight-line fits to the individual samples shown previously in Figure 9. The solid line indicates the fit to the combined sample and the dotted lines indicate the 68.3$\%$ confidence regions of this fit.}
\label{MassRichness}
\end{figure}

\begin{figure}[h!]
\epsscale{1.0}
\includegraphics[angle=-90,scale=.625]{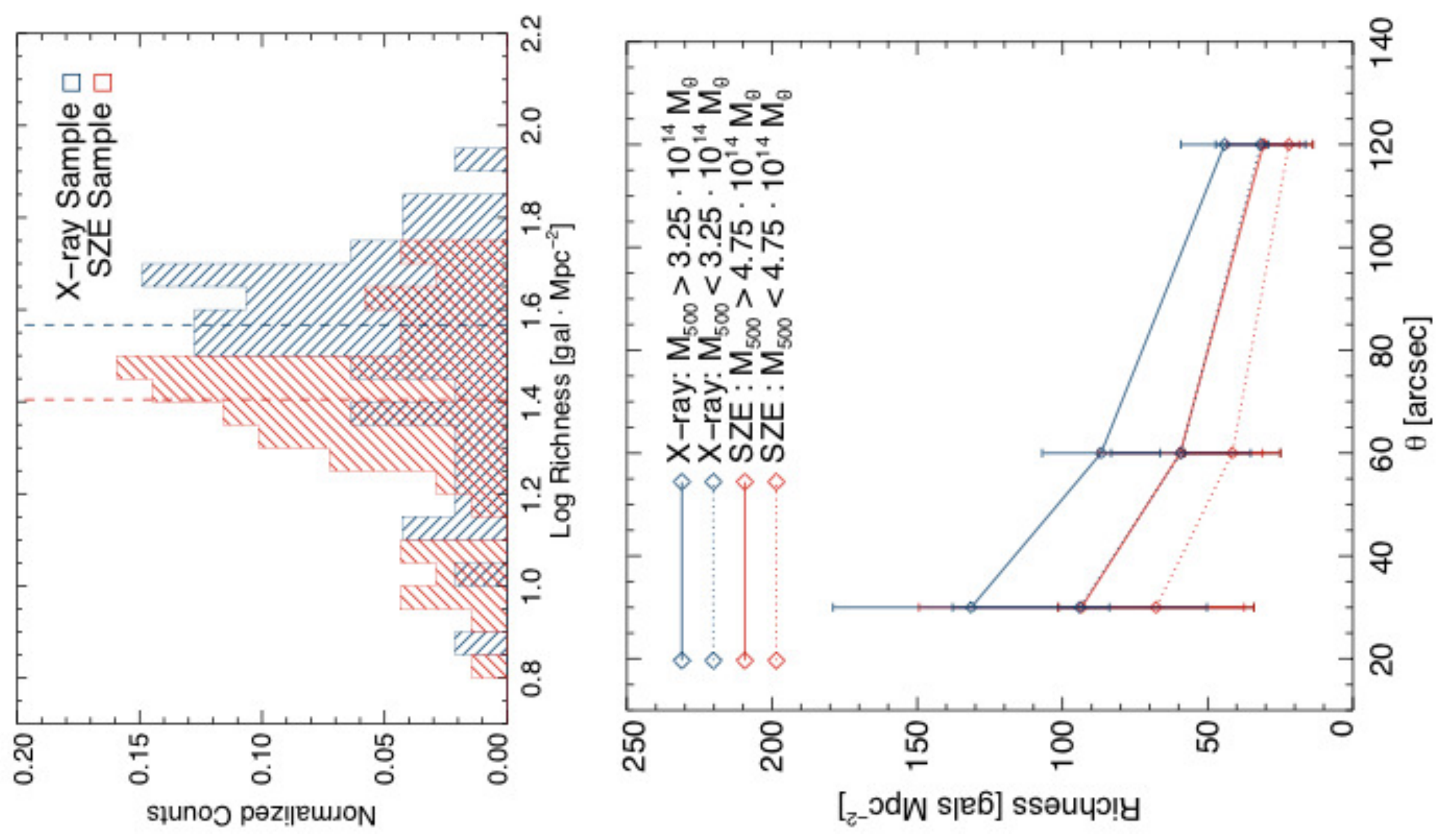}
\caption{{\it Top:} Histogram of richness of the X-ray (blue) and the SZE (red) samples. The dashed lines indicate the median richness of each sample. {\it Bottom:} The dependance of richness with cluster mass and  aperture radius  for the  X-ray (blue) and the SZE (red) samples.}
\label{Profile}
\end{figure}

\section{Calibrating a Cluster Mass-Richness relation at $0.4<z<2.0$}

In this section, we calibrate mass-richness relations based on our richness estimates defined in \S 4 and on total cluster masses as described in \S 2.1 and \S 2.2. In Fig.~\ref{MassRichnessXvsSZ} we show the relations we find for the 47 clusters of the X-ray sample (top panel) and the 69 clusters of the SZE sample (bottom panel).  We perform a least squares fit of the data for each sample individually with a single linear relation on log-quantities, where also errors in both variables are taken into account, and find:

\begin{equation}
\log \frac{M_{500,X}}{M_{\odot}}= (13.68 \pm 0.17) + (0.57 \pm 0.23) \cdot \log \frac{R_{[4.5]}}{gals \cdot Mpc^{-2}},
\label{richXequation}
\end{equation}

and

\begin{equation}
\log \frac{M_{500,SZ}}{M_{\odot}}= (13.34 \pm 0.21) + (0.93 \pm 0.22) \cdot \log \frac{R_{[4.5]}}{gals \cdot Mpc^{-2}}
\label{richYZequation}
\end{equation}

In Fig.~\ref{MassRichness} we show the relation we find for the 116 clusters of the combined sample (solid black line):

\begin{equation}
\log \frac{M_{{500}}}{M_{\odot}}= (13.56 \pm 0.25) + (0.74 \pm 0.18) \cdot \log \frac{R_{[4.5]}}{gals \cdot Mpc^{-2}}.
\label{richAllequation}
\end{equation}

To test the robustness of our fit, we have also run a bootstrap Monte Carlo test, in which the mass-richness relation is repeatedly resampled, 
to reveal whether a small sample of clusters could for instance dramatically alter the result of the fit. We run the least-square fitting algorithm 1000 times, and at each repetition we randomly toss out 25\% of the sample. We then infer the mean and standard deviation of the intercept and slope distribution for the 1000 fit results. We find the latter values in perfect agreement with the ones of Eq. ~\ref{richAllequation}.\\
We have also checked whether a small subsample of high mass clusters could largely alter the result of the fit. However, even excluding the three most massive clusters in the sample, the resulting values of intercept and slope and their errors are still consistent within 1$\sigma$ of the ones presented in Eq. ~\ref{richAllequation}.

Based on the 68.3 \% confidence regions of the fits (dotted lines), we estimate the associated errors in mass at fixed richness to be $\pm 0.25$ dex. We will discuss the dependance of the scatter of this relation with concentration in \S 6.2. The intrinsic scatter of the relation is measured in the $R_{[4.5]}$ direction around the best-fitting  $R_{[4.5]}-M$ relation for that sample, and is denoted as $\sigma_{R_{[4.5]} | M}$. We find $\sigma_{R_{[4.5]} | M}$=0.32 dex for our sample.

We compare the measured scatter in our relation with literature richness and mass estimates. Using an $r$-band luminosity-based optical richness estimator, $R_{L}$,  \citet{Planck14} found the associated error in mass at fixed richness to be $\pm 0.27$ dex. The intrinsic scatter of their $R_{L}-M$ relation, $\sigma_{R_{L} | M}$=0.35, is also similar to the value that we have found.   We remind that  $R_{L}$ was defined by \citet{Wen12} for a large sample of low-redshift Sloan Digital Sky Survey \citep[SDSS,][]{York00} selected clusters for which X-ray masses were provided by \citet{Piffaretti11}. We note that their method cannot be extended to all clusters in our sample because SDSS lacks coverage of the Southern Sky and because deeper optical data than the one available from SDSS would be required to detect the bulk of cluster galaxies at $0.6 < z \lesssim 2$.

Also at lower redshifts,  $0.03 <z< 0.55$, \citet{Andreon10} and \citet{Andreon15} have used multi-band SDSS photometry to define an optical richness estimator, $n_{200}$, aimed at counting red cluster members within a specified luminosity range and color as a proxy of the total mass, $M_{200}$,  within the $R_{200}$ radius. These predicted masses are found to have a smaller 0.16 dex scatter with mass, but require optical photometry in at least two bands.

As a comparison with other observable-mass scaling relations, we note that \citet{Maughan07} have studied the  $L_{X}-M$ relation for a sample of 115 clusters at $0< z< 1.3$, and found the associated error in mass at fixed luminosity to be $\pm 0.21$ dex. They have measured the intrinsic scatter in the $L_{X}-M$ relation to be $\sigma_{L | M}$= 0.39 when all of the core emission is included in their $L_{X}$ measurements. Interestingly, they have also demonstrated that the scatter can greatly be reduced to $\sigma_{L | M}$= 0.17 by excising the core emission in their $L_{X}$  measurements. 
Furthermore \citet{Rozo14a, Rozo14b} have used SZE data from {\it Planck} and X-ray data from {\it Chandra} and {\it XMM-Newton} to calibrate $Y_{X}-M$,  $Y_{SZ}-Y_{X}$  and $Y_{SZ}-M$ relations for different samples of clusters at $z<0.3$. They found low values for the scatters of the $Y_{SZ}-M$ relations,  $\sigma_{Y_{SZ} | M}$= 0.12-0.20.

To summarize, the method we have proposed here requires only shallow, single pointing, single  band observations and an estimate of the cluster center position and redshift to provide reliable richness-based cluster mass estimates. The very low observational cost associated with our approach  makes it potentially available for very large samples of clusters at $0.4 < z < 2.0$. 

\begin{figure}[t!]
\epsscale{1.0}
\includegraphics[scale=.4]{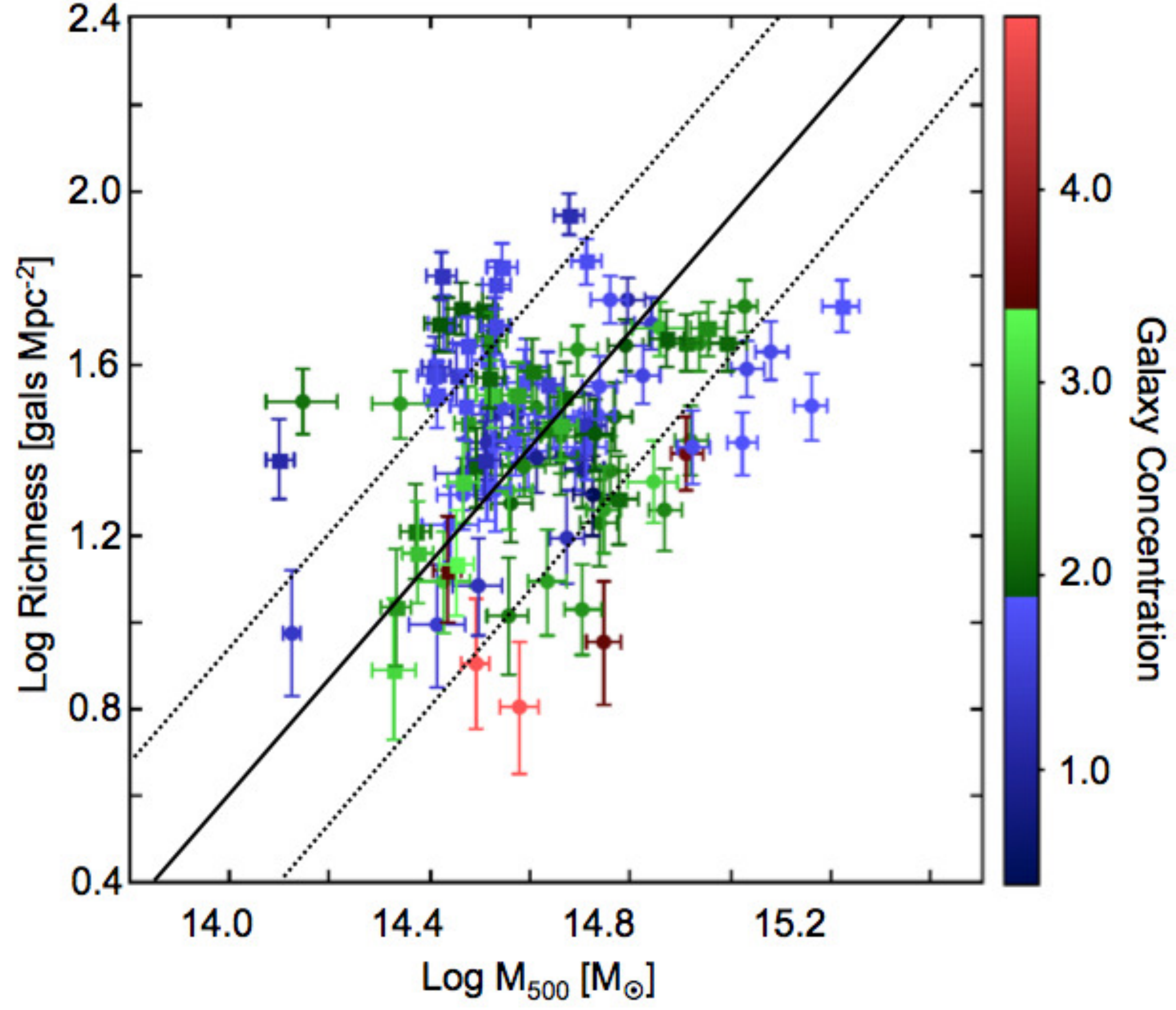}
\caption{The 4.5 $\mu$m richness vs. mass relation color-coded by galaxy concentration for the combined sample. The solid line indicates the linear fit to the sample where also errors in both variables are taken into account. The dotted lines indicate the 68.3$\%$ confidence regions of this fit.}
\label{Concentration}
\end{figure}

\begin{figure*}[t!]
\epsscale{1.0}
\includegraphics[scale=.375]{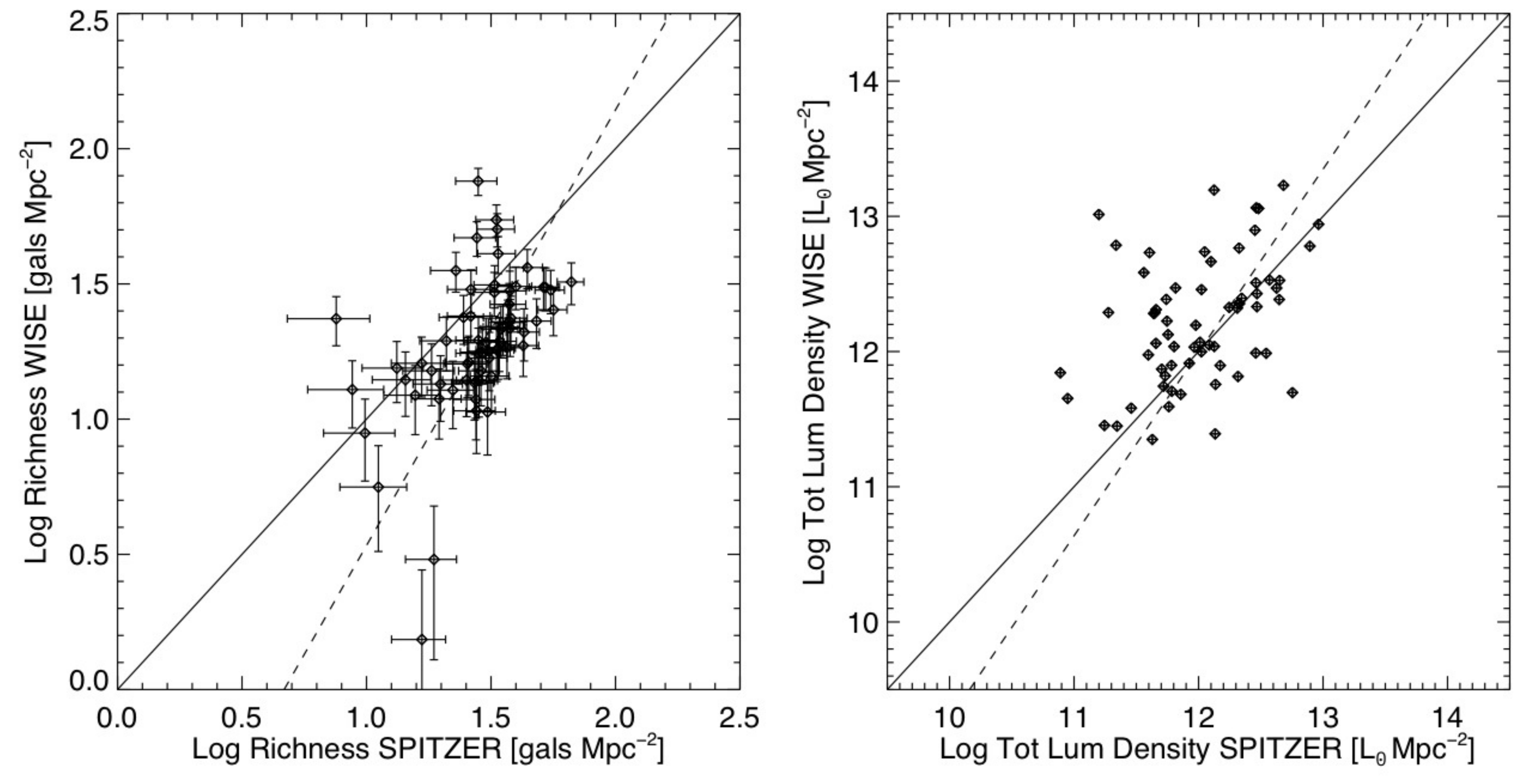}
\caption{Comparison of richness estimates (left panel) and total luminosity density of all sources measured within $r=120"$ from each cluster center (right panel) based on data from {\it Spitzer} at 4.5$\mu$m and {\it WISE} at 4.6$\mu$m. The solid lines represent the identity (1:1) lines. The dashed lines correspond to the best straight-line fit to the data with errors in both coordinates.}
\label{wise}
\end{figure*}

\section{Discussion}

In this section we discuss the richness distribution and galaxy surface density profile for our samples. We also examine potential sources of the scatter in the mass-richness relation, including sample selection and galaxy concentration. We then explore the possibility of extending our method  to other MIR all-sky surveys like {\it WISE}, and the implication of our findings on future, wide-field near-infrared cluster surveys like {\it Euclid}. 

\subsection{Richness Distribution and Profiles}

In the top panel of Fig.~\ref{Profile} we show the distributions of richness for our samples. The X-ray sample (blue histogram and blue dashed line) shows larger median richness values than the SZ sample  (red histogram and red dashed line). This difference is statistically significant as shown by the Kolmogorov-Smirnov (KS) test that provides a probability $P_{KS} \sim 10^{-6}$ that the observed distributions of richness are extracted from the same parent population. 

In the bottom panel of Fig.~\ref{Profile} we show the dependance of richness on aperture radius for both cluster samples. We divide each sample in two equally populated bins of X-ray or SZE derived cluster masses. We calculate the richness profiles by measuring the background-subtracted projected surface density of galaxies with $[4.5]<21$ AB  within $r= 30", 60" , 120"$ from the cluster center.
We find the shapes of the richness profiles to be similar for both samples and in each mass bins, and the richness values consistent within the large scatter. However, the low-mass X-ray sample appears to be as rich as the high-mass SZ sample. Hence at fixed cluster mass,  X-ray clusters appear to have, on average, higher near-infrared richness than SZ-selected clusters. 

One possibility is that this difference is due to systematics in derived masses for the X-ray and SZ samples. For example, if despite the careful re-calibration in \S 2, the X-ray masses are underestimates, then the higher richness of the X-ray clusters would imply that they are more massive clusters. Given the extensive work on calibrating X-ray and SZ observables with mass, this seems unlikely.
Yet, without having SZ, X-ray and richness estimates of the same clusters in this redshift range, we cannot definitively rule it out.

Alternately, could this be due to a selection effect, where the X-ray surveys, at fixed cluster mass, typically selects richer systems where galaxy merging has been, on average, less effective than in SZE selected clusters? X-ray surveys are indeed usually considered biased toward selecting more relaxed, more evolved systems \citep[e.g.,][]{Eckert11}.  This is because of the presence of a  surface brightness peak in the so-called `cool core' clusters. The clear peak of  X-ray emission is more easily detected in the wide shallow surveys of ROSAT, for instance, and it is considered to be associated with a decrease in the gas temperature,  hence typical of relaxed structures. 

On the other hand, most of the clusters newly discovered by {\it Planck} via SZE show clear indication of morphological disturbances in their X-ray images, suggesting a more  active dynamical state \citep{PlanckIX11}. Interestingly, \citet{Rossetti16}, measuring the offset between the X-ray peak and the BCG population, a known indicator of an active cluster dynamical state  \citep[e.g.,][]{Sanderson09, MannEbeling12}, have found evidence of the dynamical state of SZ-selected clusters to be significantly different from X-ray-selected samples, with a higher fraction of non-relaxed, merging systems.

In the hierarchical cluster formation scenario, clusters form by the infall of less massive groups along the filaments. 
Therefore while it is possible that a larger fraction of SZ-selected clusters in our sample are in a less developed stage of cluster formation, if the difference in richness between our samples was to be explained with the fact that X-ray clusters were more evolved and had accreted more galaxies within $R_{500}$, then they should also be the most massive. This has however not been found, as clearly shown in Fig.~\ref{sample} unless there are mass calibration uncertainties larger than those discussed.

An intriguing possibility is that there are differences in the intrinsic baryon fraction within $R_{500}$ relative to the cosmological baryon fraction between the cluster samples. 
If X-ray clusters at these redshifts harbor a larger baryon fraction per unit dark matter halo mass compared to SZE cluster samples within the aperture radius,
it could account for lower total masses, more efficient cooling of the intracluster medium by thermal bremsstrahlung and the resultant increased star-formation efficiency could result in a larger population of luminous galaxies translating to a higher richness. Indeed, simulations like the Millennium simulations show a factor of two variation in the baryon fraction of $>10^{14}$\,M$_{\sun}$ dark matter halos which could easily account for both the differences in derived masses and richness values. Even among the X-ray cluster sample, \citet{Vikhlinin09} show that the baryon fraction increases with increasing mass among their which is likely the origin of
a richness-mass correlation that we derive. A comparison between the density profiles of SZ clusters and X-ray clusters in \citet{PlanckIX11}, shows that SZ clusters show shallower density profiles
than X-ray selected clusters which may again argue that the baryons in SZE selected clusters are predominantly at larger radii than in the X-ray sample.
However, extracting effects such as these from the data would again require SZ, X-ray and richness estimates of the same clusters in this redshift range
while our study has rather heterogeneous samples with different origins that we have attempted to place on the same calibration scale. Apart from stating these possibilities, it is challenging for us
to definitively claim one of them as an origin for the observed difference.

\subsection{Galaxy Concentration}

In an attempt to understand the source of the scatter in the mass-richness relation, we also measure the galaxy concentration of our cluster samples, defined as the ratio between the richness measured within $r= 60"$ and $r=120"$. By definition, a higher value of galaxy concentration correspond to systems with a steeper galaxy surface density profile.
As shown in Fig.~\ref{Concentration}, there is a hint that clusters that deviate the most from the fitted mass-richness relation are also the ones with the most centrally concentrated
galaxy surface density profile. 
If we would include a correction for galaxy concentration, the scatter of the mass-richness relation would slightly decrease so that the associated error in mass at fixed richness  is found to be $\pm 0.22$ dex
indicating that surface density concentration does play a significant role in the scatter.

Possible origins for this are blending and source confusion in the IRAC image, beam dilution when the 2$\arcmin$ aperture is much larger than the cluster overdensity
or the impact of galaxy merging on richness estimates.
Images of galaxy clusters that are more centrally concentrated are more likely to be affected by the blending of some cluster galaxies in the core, given the {\it Spitzer} angular resolution. This could result in underestimating their richness. The amount of confusion is linearly proportional to the surface density of galaxies. A cluster which is 4 times as concentrated in surface density would have its richness
underestimated by 0.6 dex which is the amount of offset of the red points in Fig.~\ref{Concentration} from the best fit line.

Alternatively, systems with higher central galaxy concentration have their richness estimate, calculated within a radius, $r=120"$, biased by the fact that the aperture chosen is too large, hence their richness is relatively smaller compared to less centrally concentrated systems. However, if this was purely an observational bias, we should have found a correlation between galaxy concentration and $\theta_{500}$ derived for the hot gas in the intracluster medium, a proxy of cluster size, which is not apparent.

If mergers were responsible for the scatter in the richness-mass relation, clusters of the same total mass, that might have experienced more or less merger events in their cores with respect to their outskirts, would result in lower or higher concentration measurements, and would therefore be found to have lower or higher values of richness. In the most extreme cases, as shown in Fig.~\ref{Panel_method}, galaxy clusters of the same total mass (as probed by their gas), and at the same lookback time, may have experienced very different  evolutionary processes, resulting in large differences in richness and concentration inferred from the number and location of their cluster members.

Thus, we conclude that the scatter in the richness-mass relation that we derive is likely due to a combination of source confusion and differences in evolutionary history of the clusters.

\begin{figure*}[t!]
\epsscale{1.0}
\includegraphics[scale=.5]{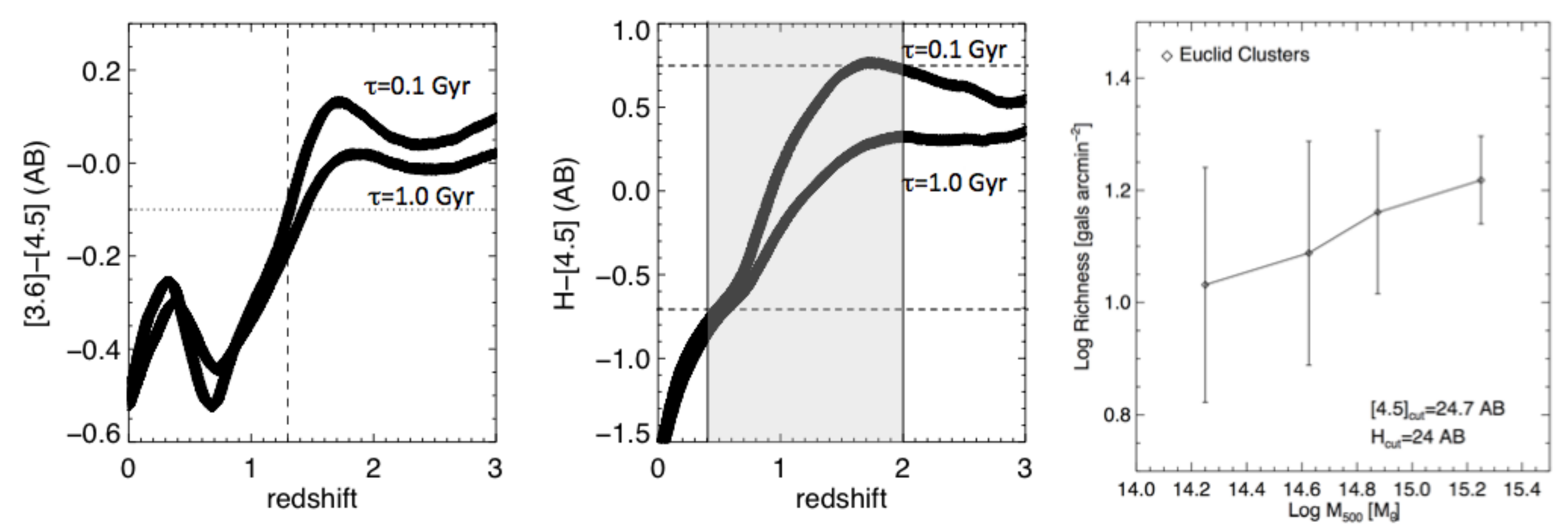}
\caption{Evolution of the $[3.6]-[4.5]$ color (left panel) and  $H-[4.5 ]$ color (middle panel) with redshift for a set of Bruzual \& Charlot (2003)  stellar population models with exponentially declining star formation rates with $\tau=0.1$ Gyr  (early-type galaxy) and $\tau=1.0$ Gyr (star forming galaxy). These colors are used to translate our measure of [4.5]\,$\mu$m richness into a H-band richness estimate. The right panel shows the predicted richness (in $gals \cdot arcmin^{-2}$) for Euclid clusters at $0.4<z<2.0$, in the wide area survey ($H_{cut}$=24 AB), as a function of cluster mass.}
\label{models}
\end{figure*}

\subsection{{\it WISE} 4.5$\mu$m Richness}

The AllWISE\footnote{\url{http://wise2.ipac.caltech.edu/docs/release/allwise/}} program combined data from the cryogenic Wide-field Infrared Survey Explorer mission \citep[{\it WISE},][]{Wright10}  and the \citep[NEOWISE][]{Mainzer11} post-cryogenic survey to deliver a survey of the full MIR sky.  Since all-sky catalogs in the W2 band at $4.6 \mu$m are available, it 
could allow us to potentially extend our method outside the {\it Spitzer} footprint. Therefore in this section we test whether our proposed method of deriving MIR richness estimates can be applied robustly to the publicly available {\it WISE} data .

To this aim, we use archival {\it WISE} 4.6$\mu$m data available at the location of all our clusters, and apply the same method described in \S 4. The SEIP Source List contains W2 photometry for positional counterparts found in the AllWISE release both for all clusters in our sample and for the control SpUDS field.  In Fig.~\ref{wise} (left panel) we show the richness estimates based on data from  {\it Spitzer} and {\it WISE} for our SZE sample. The dashed lines indicates the straight line fit to be compared to the 1:1 (solid) line. We note that there are several catastrophic outliers and that {\it Spitzer}-based richness values are systematically higher than the {\it WISE}-based counterparts.
We note that {\it WISE} is less sensitive than  {\it Spitzer} and that its angular resolution is also poorer ($6.4"$ vs. $2.0"$).

 To test whether the catastrophic discrepancies in the  {\it Spitzer} vs. {\it WISE} richness estimates could be ascribed to confusion, we sum of the flux densities of every object detected, within $r=120"$ from each cluster center, by the two instruments. We also subtract a median flux density from the background field to get a flux over density at the location of the cluster.
 Source confusion makes groups of sources in a high resolution image appear as a single bright
 source in a low resolution image. By adding the flux densities, we take out the effect of confusion which would bias richness estimates low.
 We then use the redshift of the cluster to translate the summed flux over density to a luminosity surface density.
As shown in the right panel of  Fig.~\ref{wise} we find that the total luminosity densities for each cluster appears to be conserved, as the two instruments provide matching measurements.  Therefore we can ascribe the aforementioned discrepancies solely to the poorer angular resolution of {\it WISE}, with richness estimates highly depending on the particular projected cluster galaxy geometry. At  the {\it WISE} image quality, we expect a higher number of sources to be blended, resulting in lower counts of galaxies per cluster, yielding lower richness values than those measured by {\it Spitzer}. 
For example, source over densities of 30 galaxies in the 2$\arcmin$ radius aperture would correspond to 10 gals Mpc$^{-2}$ which in turn would correspond to a ratio of 11 for the number of WISE beams
per source. This is well below the classical confusion limit. In reality, the confusion is even higher since the average underlying foreground source density would also contribute to confusion noise.

To summarize, we deem {\it WISE}-based richness estimates to be poorer proxies for cluster mass preventing us to effectively extend our method beyond the {\it Spitzer} footprint. Calibrating a Mass-richness relation for the{\it WISE} dataset will require a different technique and is therefore beyond the scope of this paper.

\subsection{Future Wide Field Near-Infrared Cluster Surveys}

The upcoming {\it Euclid} and {\it WFIRST} missions aim to survey large portions of the extra-galactic sky in the near-infrared (e.g., H band) to measure the effects of Dark Energy, but also have distant cluster studies as a key scientific goal. The {\it Euclid} wide-area survey, in particular, will observe 15000 deg$^{2}$, almost the entire extragalactic celestial sphere,  down to a 5$\sigma$ point source
depth of H=24 mag (AB).  They plan to use photometric redshift overdensities to identify clusters but that requires ancillary ground-based optical data which is currently being taken.
In this section, based on the results of our analysis,  we try to provide a simple prediction of  the expected richness values for  {\it Euclid}-selected clusters and the range of masses that will be accessible. 

{\it Euclid} is expected to detect $\sim 2 \times 10^{6}$ clusters at all redshifts, with $\sim 4 \times 10^{4}$ of them at $1 < z < 2$, with cluster masses $M_{200} \gtrsim 8 \times 10^{13} M_{\odot}$  \citep{Sartoris16}. The cluster sample in our study spans a similar mass and redshift range, hence we can attempt to predict the average richness expected for clusters at $0.4 <z < 2.0$ as a function o mass in the {\it Euclid} survey.

In Fig.~\ref{models} we show the evolution of the $[3.6]-[4.5]$ (left panel) and  $H-[4.5 ]$ color (middle panel) with redshift for a set of \citet{BC03}  stellar population models with exponentially declining star formation rates. We show both the typical color evolution expected for an early-type galaxy (assuming $\tau=0.1$ Gyr) and a star forming galaxy  ($\tau=1.0$ Gyr) as described in \citet{Rettura10,Rettura11}. As pointed out by several authors \citep[e.g.,][]{Papovich08, Muzzin13, Wylezalek13, Rettura14} the $[3.6]-[4.5]$ color is fairly insensitive to different modes of star formation out to $z\sim 3$ and can be used as a good redshift indicator at $z>1.3$.  The $H-[4.5]$ color, instead, is more sensitive to galaxy star formation history, in particular between $1 < z < 2$. 

According to the models shown in the middle panel of Fig.~\ref{models}, we expect the $H-[4.5 ]$ color of a galaxy to vary between -0.7 and 0.75 AB (dashed lines) depending on its type at $0.4 <z < 2.0$  (gray shaded area). This would imply that to match the {\it Euclid} $H_{cut}$=24 AB depth, we need an equivalent {\it Spitzer} 4.5 $\mu$m survey to reach $[4.5]_{cut}=24.7 AB$.

In \S 4.1, based on our `deep' sub-sample of 36 clusters for which the deepest IRAC coverage was available in the {\it Spitzer} archive, we have derived a relation between the survey depth and the average richness of our cluster sample. As we have demonstrated, we already are below the knee of the galaxy luminosity function at these redshifts and we do not expect the slope of the relation to change as we go deeper.
Using Eq. 3, we can then predict the richness of clusters at an equivalent depth of $[4.5]_{cut}=24.7$, i.e. down to $H_{cut}$=24. We predict the following levels of richness (in galaxies Mpc$^{-2}$) for {\it Euclid}-detected clusters as a function of cluster mass: $\log R_{[H]}=1.78 \pm 0.26$ ($\log M_{500} < 14.5$), $\log R_{[H]}=1.87 \pm 0.21$ ($14.5 \leq \log M_{500} \leq 14.75$), $\log R_{[H]}=1.99 \pm 0.16$ ($\log M_{500} > 14.75$).
In the right panel of Fig.~\ref{models} we show the expected mean richness (and standard deviation of the mean) in bins of cluster mass for the {\it Euclid} clusters at $0.4 <z < 2.0$, propagating
the uncertainty in the fit required to extrapolate to these faint flux densities. For immediate comparison with the future observational data, the figure reports the expected richness values in units of $galaxies \cdot arcmin^{-2}$.  We conclude that typical {\it Euclid} clusters that are about 3$\times10^{14}$\,M$_{\sun}$ will show galaxy over densities of $\sim$12 $galaxies \cdot arcmin^{-2}$. 

\section{Summary}

In this paper we have studied a sample of 116 X-ray and Sunyaev-Zeldovich effect selected galaxy clusters at $0.4 < z < 2.0$ observed by {\it Spitzer} at 4.5$\mu$m. Together, they span more than a decade in total cluster mass. 
With the aim of providing a simple and efficient observable that easily translates as a proxy for cluster mass, we have defined a 4.5$\mu$m richness parameter that requires just a single pointing of IRAC imaging and shallow observing time ($\sim$ 90sec) that reaches a depth of [4.5$]<$21 AB mag. 
We have obtained the following results:

\begin{itemize}
\item{} We have derived {\it ROSAT}-based X-ray bolometric luminosities and masses that are in agreement with independent studies performed using {\it Chandra} data by \citet{Maughan07, Maughan12}.
\end{itemize}

\begin{itemize}
\item{} By analyzing deeper IRAC imaging data, available for a subsample of systems, we have studied and parameterized the dependance of our richness parameter on survey depth and aperture radius. We have  found that richness measured in the larger radius adopted here, r=$2\arcmin$, is less sensitive to variations in depth.
\end{itemize}

\begin{itemize}
\item{} We have calibrated a mass-richness relation for both subsamples individually and combined. We have fitted linear relations in log-log space and estimated the associated error in mass at fixed richness to be $\pm 0.17, 0.22, 0.25$ dex
for the X-ray, SZE and the combined samples, respectively. We find a slight dependence of the scatter with galaxy concentration, defined as the ratio between richness measured within an aperture  of 1 and 2 arcminutes.
\end{itemize}

\begin{itemize}
\item{} 
We have measured the intrinsic log-scatter of our 4.5 $\mu$m richness-mass relation for our combined sample, $\sigma_{R_{[4.5]} | M}$=0.32 dex. The value of scatter we found is similar to the one obtained by the \citet{Planck14} adopting deeper SDSS-based optical richness estimator at lower redshifts. The scatter associated with our observable is larger than the one obtained by  \citet{Andreon15} and \citet{Rozo14b} that have however adopted richness estimates that require deeper  multiband observations which are challenging, particularly at high redshifts.
\end{itemize}

\begin{itemize}
\item{} We have found that similar {\it WISE}-based 4.6$\mu$m richness estimates would provide poorer proxies of cluster mass due to the lower angular resolution of the data with respect to Spitzer/IRAC that results in source confusion.
\end{itemize}

\begin{itemize}
\item{} Finally, we provide a calibration of the average richness as a function of cluster mass in the near-infrared, which can be applied to galaxy over-densities that will be detected by the upcoming {\it Euclid} mission through its wide-area near-infrared survey.
\end{itemize}

As {\it Spitzer} continues to survey large area of the sky during its extended {\it Warm Mission} our results make richness-based cluster mass estimates available for large samples of clusters at a very low observational cost up to $z \sim 2$. 

\acknowledgments  A.R.  is  grateful to the Spitzer Archival Team for providing access to advanced data products and is thankful to Dr. Peter Capak for providing access to the SEIP photometry pipeline. A.R. is grateful to Drs. M. Nonino, L. Girardi for discussions and providing access to the TRILEGAL model runs. A.R.  is  grateful to Drs. Mark Brodwin, Anthony Gonzalez,  Ben Maughan, Adam Mantz, Mauro Sereno, Loredana Vetere for interesting discussions, comments and suggestions that improved this manuscript. This work is based on data obtained with the {\it Spitzer Space Telescope}, which is operated by the Jet Propulsion Lab (JPL), California Institute of Technology (Caltech), under a contract with NASA. Support was provided by NASA through contract number 1439211 and 1484822 issued by JPL/Caltech.  

\clearpage

\begin{table} [h!]
\caption{X-ray Selected Cluster Sample}
\label{table:1} 
\centering
\tiny
\begin{tabular}{l l l l l c c c c}     % 9 columns
\hline
\hline
Clus ID	& RA	 & DEC	& NAME	& z & $\log M_{500, X}$  & 	$\log R_{[4.5]}$ & $N_{Cluster}$ &  $N_{S}$ \\
                  & (deg., J2000) & (deg., J2000) & &  & ($M\odot$) & ($galaxies \cdot Mpc^{-2}$) & & \\
\hline
\hline
  26   &  4.6408   & 16.4381  & MACS J0018.5+1626   & 0.5456 & 14.995 $\pm$  0.035  &     1.645$^{+0.061}_{-0.071}$  &       162 & 14.32 \\
  46   &  7.64     & 26.3044  & WARP J0030.5+2618   & 0.5    & 14.506 $\pm$  0.029  &     1.716$^{+0.056}_{-0.065}$  &       173 & 19.31 \\
  51   &  8.9971   & 85.2214  & WARP J0035.9+8513   & 0.8317 & 14.462 $\pm$  0.034  &     1.721$^{+0.056}_{-0.064}$  &       242 & 37.17 \\
  145  &  25.3846  & -30.5783 & 400d J0141-3034     & 0.442  & 14.514 $\pm$  0.028  &     1.604$^{+0.064}_{-0.074}$  &       134 & 8.74  \\
  156  &  28.1721  & -13.9703 & WARP J0152.7-1357   & 0.833  & 14.638 $\pm$  0.035  &     1.448$^{+0.075}_{-0.091}$  &       149 & 8.60  \\
  187  &  34.1404  & -17.7908 & WARP J0216.5-1747   & 0.578  & 14.435 $\pm$  0.030  &     1.119$^{+0.106}_{-0.140}$  &       101 & 8.75  \\
  200  &  37.6108  & 18.6061  & 400d J0230+1836     & 0.799  & 14.671 $\pm$  0.035  &     1.520$^{+0.070}_{-0.083}$  &       167 & 15.70 \\ 
  268  &  52.1504  & -21.6678 & 400d J0328-2140     & 0.59   & 14.545 $\pm$  0.031  &     1.819$^{+0.050}_{-0.057}$  &       208 & 10.59 \\
  276  &  53.2925  & -24.9447 & 400d J0333-2456     & 0.475  & 14.477 $\pm$  0.029  &     1.459$^{+0.074}_{-0.089}$  &       123 & 10.73 \\
  312  &  58.9971  & -37.6961 & 400d J0355-3741     & 0.473  & 14.528 $\pm$  0.029  &     1.683$^{+0.058}_{-0.067}$  &       155 & 12.39 \\
  316  &  61.3512  & -41.0042 & 400d J0405-4100     & 0.686  & 14.524 $\pm$  0.032  &     1.525$^{+0.069}_{-0.082}$  &       156 & 13.23 \\
  355  &  73.5462  & -3.015   & MACS J0454.1-0300   & 0.5377 & 14.954 $\pm$  0.034  &     1.676$^{+0.059}_{-0.068}$  &       178 & 25.66 \\
  380  &  80.2937  & -25.51   & 400d J0521-2530     & 0.581  & 14.491 $\pm$  0.030  &     1.357$^{+0.083}_{-0.102}$  &       135 & 23.86 \\
  382  &  80.5575  & -36.4136 & 400d J0522-3624     & 0.472  & 14.412 $\pm$  0.028  &     1.589$^{+0.065}_{-0.076}$  &       150 & 22.29 \\
  405  &  85.7117  & -41.0014 & RDCS J0542-4100     & 0.642  & 14.585 $\pm$  0.032  &     1.552$^{+0.067}_{-0.080}$  &       170 & 26.83 \\
  550  &  132.1983 & 44.9392  & RX J0848.7+4456     & 0.574  & 14.100 $\pm$  0.028  &     1.375$^{+0.081}_{-0.100}$  &       126 & 13.56 \\
  551  &  132.2346 & 44.8711  & RX J0848.9+4452     & 1.261  & 14.328 $\pm$  0.041  &     0.889$^{+0.133}_{-0.193}$  &       105 & 13.54 \\
  557  &  133.3058 & 57.9956  & 400d J0853+5759     & 0.475  & 14.435 $\pm$  0.028  &     1.683$^{+0.058}_{-0.067}$  &       157 & 14.01 \\
  586  &  141.6521 & 12.7164  & 400d J0926+1242     & 0.489  & 14.405 $\pm$  0.028  &     1.567$^{+0.066}_{-0.078}$  &       141 & 14.00 \\
  601  &  145.7796 & 46.9975  & RXC J0943.1+4659    & 0.4069 & 14.679 $\pm$  0.030  &     1.939$^{+0.044}_{-0.049}$  &       192 & 10.53 \\
  621  &  149.0121 & 41.1189  & 400d J0956+4107     & 0.587  & 14.465 $\pm$  0.030  &     1.322$^{+0.086}_{-0.107}$  &       118 & 9.88  \\
  631  &  150.5321 & 68.98    & 400d J1002+6858     & 0.5    & 14.455 $\pm$  0.029  &     1.568$^{+0.066}_{-0.078}$  &       142 & 13.35 \\
  634  &  150.7671 & 32.9078  & 400d J1003+3253     & 0.4161 & 14.711 $\pm$  0.030  &     1.830$^{+0.050}_{-0.056}$  &       168 & 9.63  \\
  713  &  164.2479 & -3.6244  & MS1054.4-0321       & 0.8309 & 14.661 $\pm$  0.035  &     1.454$^{+0.075}_{-0.090}$  &       154 & 12.61 \\
  743  &  169.375  & 17.7458  & 400d J1117+1744     & 0.547  & 14.417 $\pm$  0.029  &     1.525$^{+0.069}_{-0.082}$  &       137 & 8.73  \\
  747  &  170.0321 & 43.3019  & WARP J1120.1+4318   & 0.6    & 14.634 $\pm$  0.032  &     1.547$^{+0.068}_{-0.080}$  &       146 & 8.31  \\
  748  &  170.2429 & 23.4428  & 400d J1120+2326     & 0.562  & 14.522 $\pm$  0.030  &     1.647$^{+0.061}_{-0.071}$  &       159 & 8.37  \\
  825  &  180.5571 & 57.8647  & 400d J1202+5751     & 0.677  & 14.508 $\pm$  0.032  &     1.372$^{+0.081}_{-0.100}$  &       129 & 9.45  \\
  864  &  185.3542 & 49.3019  & 400d J1221+4918     & 0.7    & 14.605 $\pm$  0.033  &     1.580$^{+0.065}_{-0.077}$  &       163 & 8.55  \\
  865  &  185.5079 & 27.1553  & 400d J1222+2709     & 0.472  & 14.417 $\pm$  0.028  &     1.522$^{+0.069}_{-0.083}$  &       127 & 8.11  \\
  873  &  186.74   & 33.5472  & WARP J1226.9+3332   & 0.888  & 14.779 $\pm$  0.038  &     1.281$^{+0.089}_{-0.113}$  &       127 & 8.02  \\
  971  &  198.0808 & 39.0161  & 400d J1312+3900     & 0.404  & 14.426 $\pm$  0.027  &     1.688$^{+0.058}_{-0.067}$  &       139 & 8.48  \\
  1020 &  203.585  & 50.5181  & ZwCl 1332.8+5043    & 0.62   & 14.530 $\pm$  0.031  &     1.682$^{+0.058}_{-0.068}$  &       176 & 9.29  \\
  1050 &  206.875  & -11.7489 & RXC J1347.5-1144    & 0.4516 & 15.221 $\pm$  0.037  &     1.726$^{+0.056}_{-0.064}$  &       162 & 15.81 \\
  1063 &  208.57   & -2.3628  & 400d J1354-0221     & 0.546  & 14.418 $\pm$  0.029  &     1.687$^{+0.058}_{-0.067}$  &       169 & 12.94 \\ 
  1066 &  209.3308 & 62.545   & 400d J1357+6232     & 0.525  & 14.474 $\pm$  0.029  &     1.635$^{+0.061}_{-0.072}$  &       154 & 11.18 \\ 
  1089 &  213.7962 & 36.2008  & WARP J1415.1+3612   & 0.7    & 14.473 $\pm$  0.032  &     1.498$^{+0.071}_{-0.085}$  &       149 & 9.61  \\
  1094 &  214.1171 & 44.7772  & NSCS J141623+444558 & 0.4    & 14.531 $\pm$  0.028  &     1.778$^{+0.053}_{-0.060}$  &       154 & 9.77  \\
  1107 &  215.9492 & 24.0781  & MACS J1423.8+2404   & 0.543  & 14.909 $\pm$  0.034  &     1.642$^{+0.061}_{-0.071}$  &       157 & 10.25 \\
  1171 &  229.4829 & 31.4597  & WARP J1517.9+3127   & 0.744  & 14.332 $\pm$  0.032  &     1.033$^{+0.115}_{-0.158}$  &       105 & 12.11 \\
  1184 &  231.1679 & 9.9597   & WARP J1524.6+0957   & 0.516  & 14.576 $\pm$  0.030  &     1.522$^{+0.069}_{-0.083}$  &       140 & 15.69 \\
  1264 &  250.4679 & 40.0247  & 400d J1641+4001     & 0.464  & 14.520 $\pm$  0.029  &     1.564$^{+0.066}_{-0.078}$  &       142 & 18.84 \\
  1410 &  275.4087 & 68.4644  & RX J1821.6+6827     & 0.8156 & 14.453 $\pm$  0.034  &     1.135$^{+0.104}_{-0.137}$  &       132 & 29.93 \\
  1506 &  309.6225 & -1.4214  & RX J2038.4-0125     & 0.673  & 14.373 $\pm$  0.031  &     1.208$^{+0.096}_{-0.124}$  &       165 & 62.22 \\ 
  1519 &  314.0908 & -4.6308  & MS2053.7-0449       & 0.583  & 14.425 $\pm$  0.030  &     1.799$^{+0.052}_{-0.058}$  &       231 & 40.92 \\
  1548 &  322.3579 & -7.6917  & MACS J2129.4-0741   & 0.594  & 14.873 $\pm$  0.034  &     1.652$^{+0.060}_{-0.070}$  &       181 & 24.79 \\
  1658 &  345.7004 & 8.7306   & WARP J2302.8+0843   & 0.722  & 14.377 $\pm$  0.031  &     1.158$^{+0.102}_{-0.133}$  &       117 & 16.19 \\\hline
\end{tabular}
\end{table}

\clearpage

\begin{table}[h!]
\caption{SZE Selected Cluster Sample}
\label{table:2} 
\centering
\tiny
\begin{tabular}{l l l l l c c c c} % 9 columns
\hline
\hline
Clus ID	& RA	 & DEC	& NAME	& z & $\log M_{500, SZ}$  & 	$\log R_{[4.5]}$ & $N_{Cluster}$ &  $N_{S}$ \\
                  & (deg., J2000) & (deg., J2000) & &  & ($M\odot$) & ($galaxies \cdot Mpc^{-2}$) & & \\
\hline
\hline
 OBJ1  & 3.05417    & 16.0375    &  MOO J0012+1602$^{(1)}$         & 0.944  & 14.146$^{+0.071}_{-0.085}$ & 1.505$^{+0.069}_{-0.082}$ &   172 & 14.39 \\
 OBJ4  & 49.8517    & -0.4225    &  MOO J0319-0025         & 1.194  & 14.491$^{+0.027}_{-0.029}$ & 0.902$^{+0.126}_{-0.178}$ &   104 & 12.19 \\
 OBJ5  & 153.535004 & 0.64056    &  MOO J1014+0038         & 1.27   & 14.531$^{+0.025}_{-0.026}$ & 1.423$^{+0.075}_{-0.091}$ &   165 & 13.49 \\
 OBJ7  & 228.677917 & 13.77528   &  MOO J1514+1346         & 1.059  & 14.342$^{+0.055}_{-0.064}$ & 1.501$^{+0.069}_{-0.083}$ &   176 & 14.02 \\
 OBJ8  & 216.637299 & 35.139889  &  IDCS J1426.5+3508$^{(2)}$     & 1.75   & 14.415$^{+0.055}_{-0.063}$ & 0.990$^{+0.120}_{-0.166}$ &   109 & 9.94  \\
 OBJ10 & 34.432999  & -3.76      &  XLSSU J021744.1-034536$^{(3)}$ & 1.91   & 14.127$^{+0.017}_{-0.018}$ & 0.973$^{+0.122}_{-0.171}$ &   107 & 9.52  \\
 OBJ9  & 86.655128  & -53.757099 &  SPT-CL J0546-5345$^{(4)}$      & 1.067  & 14.703$^{+0.034}_{-0.037}$ & 1.346$^{+0.075}_{-0.091}$ &   161 & 27.47 \\
 OBJ11 & 310.248322 & -44.860229 &  SPT-CL J2040-4451      & 1.478  & 14.522$^{+0.041}_{-0.045}$ & 1.395$^{+0.072}_{-0.087}$ &   177 & 28.46 \\
 OBJ12 & 31.442823  & -58.48521  &  SPT-CL J0205-5829      & 1.322  & 14.675$^{+0.034}_{-0.037}$ & 1.194$^{+0.095}_{-0.122}$ &   130 & 12.67 \\
 OBJ13 & 316.52063  & -58.745075 &  SPT-CL J2106-5844      & 1.132  & 14.922$^{+0.031}_{-0.033}$ & 1.418$^{+0.072}_{-0.086}$ &   172 & 24.45 \\
 OBJ16 & 355.299103 & -51.328072 &  SPT-CL J2341-5119      & 1.003  & 14.747$^{+0.033}_{-0.036}$ & 0.950$^{+0.120}_{-0.166}$ &   105 & 12.16 \\
 OBJ17 & 93.964989  & -57.776272 &  SPT-CL J0615-5746      & 0.972  & 15.023$^{+0.031}_{-0.033}$ & 1.410$^{+0.068}_{-0.081}$ &   176 & 35.12 \\
 OBJ18 & 326.64624  & -46.550034 &  SPT-CL J2146-4633      & 0.933  & 14.737$^{+0.035}_{-0.038}$ & 1.228$^{+0.088}_{-0.110}$ &   132 & 17.61 \\
 OBJ20 & 83.400879  & -50.09008  &  SPT-CL J0533-5005      & 0.881  & 14.578$^{+0.040}_{-0.044}$ & 0.802$^{+0.126}_{-0.179}$ &   100 & 24.49 \\
 OBJ21 & 15.729427  & -49.26107  &  SPT-CL J0102-4915      & 0.8701 & 15.159$^{+0.030}_{-0.033}$ & 1.496$^{+0.071}_{-0.085}$ &   162 & 10.60 \\
 OBJ22 & 9.175811   & -44.184902 &  SPT-CL J0036-4411      & 0.869  & 14.512$^{+0.047}_{-0.052}$ & 1.414$^{+0.077}_{-0.094}$ &   147 & 10.13 \\
 OBJ23 & 72.27417   & -49.024605 &  SPT-CL J0449-4901      & 0.792  & 14.69 $^{+0.036}_{-0.039}$ & 1.384$^{+0.076}_{-0.092}$ &   146 & 17.81 \\
 OBJ24 & 359.922974 & -50.164902 &  SPT-CL J2359-5009      & 0.775  & 14.557$^{+0.041}_{-0.045}$ & 1.013$^{+0.114}_{-0.154}$ &   104 & 11.54 \\
 OBJ25 & 353.105713 & -53.967545 &  SPT-CL J2332-5358      & 0.402  & 14.723$^{+0.036}_{-0.039}$ & 1.291$^{+0.083}_{-0.103}$ &   105 & 12.88 \\
 OBJ26 & 325.139099 & -57.457577 &  SPT-CL J2140-5727      & 0.4054 & 14.531$^{+0.048}_{-0.054}$ & 1.476$^{+0.065}_{-0.077}$ &   126 & 20.02 \\
 OBJ27 & 69.574867  & -54.321243 &  SPT-CL J0438-5419      & 0.4214 & 15.033$^{+0.031}_{-0.034}$ & 1.582$^{+0.061}_{-0.071}$ &   137 & 17.77 \\
 OBJ28 & 87.904144  & -57.155659 &  SPT-CL J0551-5709      & 0.423  & 14.696$^{+0.037}_{-0.041}$ & 1.626$^{+0.054}_{-0.062}$ &   154 & 28.82 \\
 OBJ29 & 62.815441  & -48.321751 &  SPT-CL J0411-4819      & 0.4235 & 14.913$^{+0.032}_{-0.035}$ & 1.387$^{+0.075}_{-0.091}$ &   115 & 14.52 \\
 OBJ30 & 323.916351 & -57.44091  &  SPT-CL J2135-5726      & 0.427  & 14.789$^{+0.034}_{-0.037}$ & 1.638$^{+0.057}_{-0.065}$ &   148 & 20.49 \\
 OBJ31 & 321.146179 & -61.410179 &  SPT-CL J2124-6124      & 0.435  & 14.715$^{+0.037}_{-0.040}$ & 1.453$^{+0.065}_{-0.077}$ &   130 & 22.74 \\
 OBJ32 & 52.728668  & -52.469772 &  SPT-CL J0330-5228      & 0.4417 & 14.824$^{+0.034}_{-0.036}$ & 1.570$^{+0.064}_{-0.075}$ &   134 & 13.24 \\
 OBJ33 & 77.337387  & -53.705322 &  SPT-CL J0509-5342      & 0.4607 & 14.704$^{+0.037}_{-0.040}$ & 1.028$^{+0.091}_{-0.116}$ &   104 & 21.00 \\
 OBJ34 & 60.968086  & -57.323669 &  SPT-CL J0403-5719      & 0.4664 & 14.574$^{+0.044}_{-0.049}$ & 1.475$^{+0.069}_{-0.081}$ &   129 & 15.99 \\
 OBJ35 & 103.962601 & -52.567741 &  SPT-CL J0655-5234      & 0.4703 & 14.707$^{+0.038}_{-0.042}$ & 1.351$^{+0.056}_{-0.065}$ &   158 & 56.19 \\
 OBJ36 & 326.468201 & -56.747559 &  SPT-CL J2145-5644      & 0.48   & 14.840$^{+0.033}_{-0.036}$ & 1.688$^{+0.055}_{-0.063}$ &   164 & 19.35 \\
 OBJ37 & 308.801147 & -52.851883 &  SPT-CL J2035-5251      & 0.5279 & 14.793$^{+0.035}_{-0.038}$ & 1.741$^{+0.050}_{-0.057}$ &   194 & 29.60 \\
 OBJ38 & 354.352264 & -59.704929 &  SPT-CL J2337-5942      & 0.775  & 14.926$^{+0.031}_{-0.034}$ & 1.402$^{+0.076}_{-0.092}$ &   144 & 14.14 \\
 OBJ39 & 82.019592  & -53.002384 &  SPT-CL J0528-5300      & 0.7678 & 14.562$^{+0.041}_{-0.046}$ & 1.273$^{+0.080}_{-0.098}$ &   137 & 23.77 \\
 OBJ40 & 345.466888 & -55.776756 &  SPT-CL J2301-5546      & 0.748  & 14.429$^{+0.052}_{-0.060}$ & 1.090$^{+0.102}_{-0.133}$ &   111 & 14.34 \\
 OBJ41 & 31.279436  & -64.545746 &  SPT-CL J0205-6432      & 0.744  & 14.532$^{+0.045}_{-0.050}$ & 1.303$^{+0.083}_{-0.103}$ &   130 & 14.62 \\
 OBJ42 & 310.8284   & -50.593838 &  SPT-CL J2043-5035      & 0.7234 & 14.656$^{+0.040}_{-0.043}$ & 1.474$^{+0.066}_{-0.077}$ &   165 & 27.70 \\
 OBJ43 & 314.217407 & -54.993736 &  SPT-CL J2056-5459      & 0.718  & 14.545$^{+0.043}_{-0.048}$ & 1.491$^{+0.065}_{-0.077}$ &   165 & 25.30 \\
 OBJ44 & 315.093262 & -45.805138 &  SPT-CL J2100-4548      & 0.7121 & 14.466$^{+0.057}_{-0.066}$ & 1.340$^{+0.075}_{-0.091}$ &   142 & 24.02 \\
 OBJ45 & 47.629108  & -46.783417 &  SPT-CL J0310-4647      & 0.7093 & 14.635$^{+0.040}_{-0.044}$ & 1.091$^{+0.105}_{-0.140}$ &   107 & 11.51 \\
 OBJ46 & 19.598965  & -51.943447 &  SPT-CL J0118-5156      & 0.705  & 14.575$^{+0.044}_{-0.049}$ & 1.449$^{+0.074}_{-0.090}$ &   143 & 11.00 \\
 OBJ47 & 0.249912   & -57.806423 &  SPT-CL J0000-5748      & 0.7019 & 14.659$^{+0.037}_{-0.040}$ & 1.378$^{+0.078}_{-0.096}$ &   135 & 13.07 \\
 OBJ48 & 68.254105  & -56.502499 &  SPT-CL J0433-5630      & 0.692  & 14.496$^{+0.050}_{-0.056}$ & 1.080$^{+0.098}_{-0.127}$ &   112 & 17.78 \\
 OBJ49 & 80.301186  & -51.076565 &  SPT-CL J0521-5104      & 0.6755 & 14.614$^{+0.039}_{-0.043}$ & 1.494$^{+0.066}_{-0.078}$ &   159 & 22.39 \\
 OBJ50 & 38.255245  & -58.327393 &  SPT-CL J0233-5819      & 0.663  & 14.594$^{+0.041}_{-0.046}$ & 1.391$^{+0.077}_{-0.094}$ &   134 & 13.00 \\
 OBJ51 & 33.106094  & -46.950199 &  SPT-CL J0212-4657      & 0.6553 & 14.770$^{+0.034}_{-0.038}$ & 1.473$^{+0.073}_{-0.087}$ &   142 & 10.42 \\
 OBJ52 & 335.712189 & -48.573456 &  SPT-CL J2222-4834      & 0.6521 & 14.734$^{+0.036}_{-0.039}$ & 1.477$^{+0.070}_{-0.084}$ &   147 & 15.01 \\
 OBJ53 & 77.920914  & -51.904373 &  SPT-CL J0511-5154      & 0.645  & 14.611$^{+0.040}_{-0.044}$ & 1.376$^{+0.074}_{-0.089}$ &   139 & 21.06 \\
 OBJ54 & 85.716667  & -41.004444 &  SPT-CL J0542-4100      & 0.642  & 14.713$^{+0.038}_{-0.041}$ & 1.405$^{+0.069}_{-0.082}$ &   148 & 26.83 \\
 OBJ55 & 40.861546  & -59.512436 &  SPT-CL J0243-5930      & 0.6352 & 14.661$^{+0.038}_{-0.042}$ & 1.427$^{+0.074}_{-0.090}$ &   137 & 13.56 \\
 OBJ56 & 319.731659 & -50.932484 &  SPT-CL J2118-5055      & 0.6254 & 14.557$^{+0.047}_{-0.053}$ & 1.301$^{+0.078}_{-0.095}$ &   130 & 21.38 \\
 OBJ57 & 326.531036 & -48.780003 &  SPT-CL J2146-4846      & 0.623  & 14.592$^{+0.044}_{-0.049}$ & 1.585$^{+0.062}_{-0.072}$ &   165 & 17.95 \\
 OBJ58 & 89.925095  & -52.826031 &  SPT-CL J0559-5249      & 0.609  & 14.762$^{+0.034}_{-0.037}$ & 1.349$^{+0.070}_{-0.083}$ &   143 & 30.61 \\
 OBJ59 & 314.587891 & -56.14529  &  SPT-CL J2058-5608      & 0.606  & 14.468$^{+0.053}_{-0.060}$ & 1.294$^{+0.076}_{-0.092}$ &   132 & 25.23 \\
 OBJ60 & 326.69574  & -57.614769 &  SPT-CL J2146-5736      & 0.6022 & 14.570$^{+0.043}_{-0.047}$ & 1.417$^{+0.072}_{-0.086}$ &   139 & 19.48 \\
 OBJ61 & 356.184692 & -42.720924 &  SPT-CL J2344-4243      & 0.596  & 15.081$^{+0.031}_{-0.033}$ & 1.625$^{+0.062}_{-0.072}$ &   162 & 10.91 \\
 OBJ62 & 64.345047  & -47.813923 &  SPT-CL J0417-4748      & 0.581  & 14.870$^{+0.033}_{-0.035}$ & 1.258$^{+0.086}_{-0.107}$ &   117 & 14.87 \\
 OBJ63 & 83.608215  & -59.625652 &  SPT-CL J0534-5937      & 0.5761 & 14.439$^{+0.055}_{-0.064}$ & 1.223$^{+0.079}_{-0.096}$ &   125 & 25.93 \\
 OBJ64 & 352.960846 & -50.863926 &  SPT-CL J2331-5051      & 0.576  & 14.748$^{+0.034}_{-0.037}$ & 1.256$^{+0.088}_{-0.111}$ &   114 & 12.34 \\
 OBJ65 & 327.181213 & -61.277969 &  SPT-CL J2148-6116      & 0.571  & 14.649$^{+0.039}_{-0.043}$ & 1.473$^{+0.067}_{-0.080}$ &   144 & 20.27 \\
 OBJ66 & 74.116264  & -51.27684  &  SPT-CL J0456-5116      & 0.5615 & 14.707$^{+0.036}_{-0.040}$ & 1.450$^{+0.069}_{-0.083}$ &   139 & 19.00 \\
 OBJ67 & 38.187614  & -52.957821 &  SPT-CL J0232-5257      & 0.5559 & 14.729$^{+0.036}_{-0.040}$ & 1.432$^{+0.075}_{-0.090}$ &   129 & 11.75 \\
 OBJ68 & 305.027344 & -63.243397 &  SPT-CL J2020-6314      & 0.5361 & 14.515$^{+0.048}_{-0.054}$ & 1.306$^{+0.069}_{-0.082}$ &   137 & 33.81 \\
 OBJ69 & 304.483551 & -62.978218 &  SPT-CL J2017-6258      & 0.5346 & 14.587$^{+0.042}_{-0.047}$ & 1.359$^{+0.066}_{-0.078}$ &   142 & 34.27 \\
 OBJ70 & 346.729767 & -65.091042 &  SPT-CL J2306-6505      & 0.5298 & 14.758$^{+0.036}_{-0.039}$ & 1.742$^{+0.053}_{-0.060}$ &   182 & 16.94 \\
 OBJ71 & 56.724724  & -54.650532 &  SPT-CL J0346-5439      & 0.5297 & 14.738$^{+0.036}_{-0.039}$ & 1.541$^{+0.066}_{-0.077}$ &   143 & 14.41 \\
 26    & 4.640833   & 16.438056  &  MACS J0018.5+1626$^{(5)}$      & 0.5456 & 14.938$^{+0.040}_{-0.044}$ & 1.645$^{+0.061}_{-0.071}$ &   162 & 14.31 \\
 355   & 73.54625   & -3.015     &  MACS J0454.1-0300$^{(5)}$      & 0.5377 & 14.858$^{+0.054}_{-0.061}$ & 1.676$^{+0.059}_{-0.068}$ &   178 & 25.66 \\
 621   & 149.012083 & 41.118889  &  400d J0956+4107$^{(5)}$        & 0.587  & 14.844$^{+0.049}_{-0.055}$ & 1.322$^{+0.086}_{-0.107}$ &   118 & 9.89  \\
 1050  & 206.875    & -11.748889 &  RXC J1347.5-1144$^{(5)}$       & 0.4516 & 15.026$^{+0.029}_{-0.031}$ & 1.726$^{+0.056}_{-0.064}$ &   162 & 15.81 \\
\hline
\end{tabular}
Note: $M_{500, SZ}$ values as reported by (1) \citet{Brodwin15}, (2) \citet{Brodwin12}, (3) \citet{Mantz14}, (4) \citet{Bleem15}, (5) \citet{Planck15}
\end{table}

\clearpage

\email{aastex-help@aas.org}

%% To help institutions obtain information on the effectiveness of their
%% telescopes, the AAS Journals has created a group of keywords for telescope
%% facilities. A common set of keywords will make these types of searches
%% significantly easier and more accurate. In addition, they will also be
%% useful in linking papers together which utilize the same telescopes
%% within the framework of the National Virtual Observatory.
%% See the AASTeX Web site at http://www.journals.uchicago.edu/AAS/AASTeX
%% for information on obtaining the facility keywords.

%% After the acknowledgments section, use the following syntax and the
%% \facility{} macro to list the keywords of facilities used in the research
%% for the paper.  Each keyword will be checked against the master list during
%% copy editing.  Individual instruments or configurations can be provided 
%% in parentheses, after the keyword, but they will not be verified.

%{\it Facilities:} \facility{Nickel}, \facility{HST (STIS)}, \facility{CXO (ASIS)}.
{\it Facilities:} \facility{Spitzer}

%% Appendix material should be preceded with a single \appendix command.
%% There should be a \section command for each appendix. Mark appendix
%% subsections with the same markup you use in the main body of the paper.

%% Each Appendix (indicated with \section) will be lettered A, B, C, etc.
%% The equation counter will reset when it encounters the \appendix
%% command and will number appendix equations (A1), (A2), etc.

\end{document}